\documentclass[twocolumn]{aastex63}

\usepackage{amsmath}
\usepackage{color}
\usepackage{graphicx}
\graphicspath{{./images/}}
\usepackage[caption=false]{subfig}
\usepackage{tikz}
\usetikzlibrary{shapes.geometric, arrows}
\usepackage{multirow}
\usepackage{enumitem}
\usepackage{algpseudocode}
\usepackage{amsmath}

\DeclareUnicodeCharacter{2212}{-}

\shorttitle{A Bayesian Investigation of Sky Models Below 6 MHz}
\shortauthors{Bassett et al.}

\begin{document}

\title{Constraining a Model of the Radio Sky Below 6 MHz Using the Parker Solar Probe/FIELDS Instrument in Preparation for Upcoming Lunar-based Experiments}

\correspondingauthor{Neil Bassett}
\author{Neil Bassett}
\affiliation{Center for Astrophysics and Space Astronomy, University of Colorado, Boulder, CO 80309, USA}
\affiliation{Department of Astrophysical and Planetary Sciences, University of Colorado, Boulder, CO 80309, USA}

\author{David Rapetti}
\affiliation{NASA Ames Research Center, Moffett Field, CA 94035, USA}
\affiliation{Research Institute for Advanced Computer Science, Universities Space Research Association, Washington, DC 20024, USA}
\affiliation{Center for Astrophysics and Space Astronomy, University of Colorado, Boulder, CO 80309, USA}

\author{Bang~D.~Nhan}
\affiliation{Central Development Laboratory (CDL), National Radio Astronomy Observatory  (NRAO), Charlottesville, VA 22903, USA}

\author{Brent Page}
\affiliation{Physics Department, University of California Berkeley, Berkeley, CA 94720, USA}
\affiliation{Space Sciences Laboratory, University of California Berkeley, Berkeley, CA 94720, USA}

\author{Jack~O.~Burns}
\affiliation{Center for Astrophysics and Space Astronomy, University of Colorado, Boulder, CO 80309, USA}
\affiliation{Department of Astrophysical and Planetary Sciences, University of Colorado, Boulder, CO 80309, USA}

\author{Marc Pulupa}
\affiliation{Space Sciences Laboratory, University of California Berkeley, Berkeley, CA 94720, USA}

\author{Stuart~D.~Bale}
\affiliation{Physics Department, University of California Berkeley, Berkeley, CA 94720, USA}
\affiliation{Space Sciences Laboratory, University of California Berkeley, Berkeley, CA 94720, USA}

\email{Neil.Bassett@colorado.edu}

\begin{abstract}

We present a Bayesian analysis of data from the FIELDS instrument on board the Parker Solar Probe (PSP) spacecraft with the aim of constraining low frequency ($\lesssim$ 6 MHz) sky in preparation for several upcoming lunar-based experiments. We utilize data recorded during PSP's ``coning roll'' maneuvers, in which the axis of the spacecraft is pointed 45$^{\circ}$ off of the Sun. The spacecraft then rotates about a line between the Sun and the spacecraft with a period of 24 minutes. We reduce the data into two formats: roll-averaged, in which the spectra are averaged over the roll, and phase-binned, in which the spectra are binned according to the phase of the roll. We construct a forward model of the FIELDS observations that includes numerical simulations of the antenna beam, an analytic emissivity function of the galaxy, and estimates of the absorption due to free electrons. Fitting 5 parameters, we find that the roll-averaged data can be fit well by this model and we obtain posterior parameter constraints that are in general agreement with previous estimates. The model is not, however, able to fit the phase-binned data well, likely due to limitations such as the lack of non-smooth emission structure at both small and large scales, enforced symmetry between the northern and southern galactic hemispheres, and large uncertainties in the free electron density. This suggests that significant improvement in the low frequency sky model is needed in order to fully and accurately represent the sky at frequencies below 6 MHz.

\end{abstract}

\keywords{cosmology: dark ages, reionization, first stars---cosmology: observations---methods: data analysis}

\section{Introduction}
\label{Introduction}

The rapidly growing field of 21 cm cosmology endeavors to open a new window into previously unobserved epochs of the early universe, starting with the Dark Ages, the period immediately following recombination, as well as Cosmic Dawn, when the first astrophysical objects coalesced, through the Epoch of Reionization (EoR). The cosmological 21 cm signal contains an abundance of information about the state and evolution of the universe during these formative periods. Making 21 cm radiation particularly useful as an observational probe is the fact that the expansion of the universe relates each redshifted frequency at which the signal can be observed today ($\sim$1-200 MHz) to a different time in the history of the universe, with lower frequencies corresponding to earlier times. However, observational measurement, particularly of the sky-averaged ``global signal,'' requires separating it from strong systematics, most notably foreground emission from both galactic and extragalactic sources.

The Dark Ages portion of the 21 cm signal (below $\sim$ 50 MHz), is particularly difficult to measure due to the increased magnitude of the foreground sky and increased ionospheric effects at lower frequencies for ground-based instruments. However, several upcoming experiments are preparing for 21 cm Dark Ages observations by performing measurements from the Moon. These experiments, Radiowave Observations at the Lunar Surface of the photoElectron Sheath (ROLSES; expected launch 2023) and the Lunar Surface Electromagnetics Experiment (LuSEE), will conduct low frequency observations that will be free of terrestrial ionospheric effects. ROLSES will be delivered to the nearside in 2023. the LuSEE experiment is composed of two separate payloads, LuSEE-lite, which will launch to the south pole farside in 2024, and LuSEE-Night, which will land at a mid-latitude farside location in 2025. All three of these experiments have been selected for flight through NASA's Commercial Lunar Payload Services (CLPS) program \citep{Burns:2021PSJ}. Although lunar observations are free of earth's ionosphere, any potential measurement of the global 21 cm signal still requires extremely careful data analysis.

Some analysis schemes for extracting the global signal from low frequency observations (such as that used by the EDGES team in \citealt{Bowman:2018}, the first reported detection of the global signal) use a linear, polynomial-based model to fit the foreground component. \cite{Tauscher:2020b} found, however, that this model of the foreground can produce false detections, particularly when used in conjunction with the flattened Gaussian signal model utilized by \cite{Bowman:2018}. The alternative is to forward-model the foreground component of observations, requiring detailed knowledge of the spatial and spectral structure of the sky (as well as the antenna beam).

Previous efforts to combine measurements and theoretical knowledge of the characteristics of the emission to produce a model of the sky at low radio frequencies include the Global Sky Model (GSM; \citealt{deOliveiraCosta:2008, Zheng:2017}) and the Ultralong-wavelength Sky Model with Absorption Effect (ULSA; \citealt{Cong:2021}). The GSM decomposes sky maps into five physical components which are then interpolated in frequency in order to output simulated maps between 10 MHz and 5 THz. A key limitation of the GSM, however, is that it does not account for free-free absorption, which becomes significant below 10 MHz, preventing the GSM from producing maps below this frequency. ULSA seeks to extend sky simulations down to 1 MHz by incorporating absorption due to free electrons within the galaxy, as well as in the circumgalactic medium (CGM) and intergalactic medium (IGM).

Limiting the accuracy of simulations of the sky below 10 MHz is the lack of sky surveys at these frequencies, even partial ones. Although a number of observations of the radio sky below 10 MHz have been attempted (see \cite{Page:2022} for a more detailed overview of previous measurements), ground-based instruments are severely limited by the ionosphere, which becomes opaque to incoming radiation at the plasma frequency ($\sim$10 MHz). Some experiments have been able to take advantage of particularly favorable ionospheric conditions in some locations to observe the emission spectrum (e.g. \citealt{Ellis:1982} and \citealt{Cane:1977} both present sky measurements below 16.5 MHz), but are derived from only a small portion of the sky. Space-based instruments have the advantage of being above Earth's ionosphere and in general have access to a larger portion of the sky due to the lack of blockage from the ground, but previous experiments have not had the capacity to perform high spatial resolution measurements as would be possible with an interferometric array. The result is that no high resolution sky surveys exist in the free-free absorption frequency regime as they do at higher frequencies (e.g. \citealt{Guzman:2011} at 45 MHz or \citealt{Haslam:1982} at 408 MHz).

We do, however, have a sense of the general characteristic of the spectrum and spatial distribution of the radio sky below 10 MHz. The ground-based measurements presented in \cite{Cane:1978} show that the brightness of the sky increases as frequency decreases until reaching $\sim$3 MHz, where the spectrum reaches a maximum before decreasing at lower frequencies\footnote{To be clear, brightness as used here refers to a quantity with units $\textup{W}\ \textup{m}^{-2}\ \textup{Hz}^{-1}\ \textup{sr}^{-1}$. The spectrum in the commonly used quantity of brightness temperature (with units K) continues to increase below 3 MHz due to the Rayleigh-Jeans law having a $\nu^{-2}$ dependence.}. In complementary space-based measurements using the WIND-WAVES instrument, \cite{Manning:2001} showed that this maximum in the spectrum near 3 MHz corresponds to a shift in the region of apparent maximum brightness from the lower galactic latitudes (above $\sim$3 MHz) to higher galactic latitudes (below $\sim$3 MHz). Both of these features are consistent with free-free absorption, which is concentrated in the plane of the galaxy and increases at lower frequencies.

In an effort to more precisely measure the radio sky at frequencies below 10 MHz, \cite{Page:2022} (hereafter referred to as P22) analyzes data from the FIELDS instrument on board Parker Solar Probe (PSP). Specifically, P22 decomposes the spectra recorded during a series of ``coning roll'' spacecraft maneuvers into spherical harmonic components. The analysis is able to constrain the spatial distribution of the emission through the $l=0$ and $l=2$ spherical harmonic functions. P22 confirms the presence of a maximum in the monopole component near 3 MHz and a shift in the maximum apparent brightness from the galactic plane to the galactic poles, consistent with the analysis of \cite{Manning:2001}, and indicates that the FIELDS antennas can be treated as ideal short dipoles in the 0.5 - 7 MHz band.

Although P22 finds that the spherical harmonic coefficients extracted from the FIELDS coning roll data are roughly consistent with synthetic sky maps in which the emissivity follows a power law spectral dependence and free-free absorption is modeled using an existing free-electron model, P22 does not attempt to fit a physical model of the sky to the data directly. Instead, the spherical harmonic decomposition presented in P22 is a model-agnostic analysis that is not dependent on any knowledge or assumptions about the emission and absorption. This paper is intended to be a companion to P22, expanding on it by utilizing the same FIELDS coning roll data, with the goal of constraining physical parameters through forward-modeling of the observations.

\section{Data}

The high frequency Receiver (HFR) of the Radio Frequency Spectrometer (RFS; \citealt{Pulupa:2017}), one component of the FIELDS instrument suite \citep{Bale:2016}, provides spectral measurements of the sky from 1.3 to 19.2 MHz. FIELDS employs four electric field sensors (V1-V4), each consisting of a 2 m long ``whip'' that is 1/8 inch in diameter, that extend perpendicularly from the PSP spacecraft in a nearly orthogonal configuration\footnote{Antenanns V1 and V2, as well as V3 and V4, are oriented 180$^{\circ}$ apart, but V1 and V3 are separated by only 85$^{\circ}$.}. The four sensors make up a crossed dipole antenna configuration with tip-to-tip lengths of 6.975 m (V1-V2) and 6.889 m (V3-V4). For a more detailed description of the FIELDS instrument and signal processing chain, see Section 2 of P22.

In order to extract any spatial information from FIELDS, we utilize FIELDS observations made at different positions relative to the Galaxy. Since PSP is pointed towards the Sun the vast majority of the time, there is very little relative motion between the FIELDS antennas and the sky. Although a Sun-pointing vector will move relative to a galactic coordinate system over the period of a PSP orbit, analysis is complicated by the changing plasma environment and its effect on quasi-thermal noise (QTN; \citealt{Meyer-Vernet:2017}) in the RFS band as PSP moves closer or further from the Sun. However, the coning roll maneuvers performed by PSP at solar distances of approximately 0.8 AU provide an ideal opportunity where the pointing of the FIELDS antennas move significantly relative to a galactic coordinate system in a regular and predictable manner. During a coning roll, the PSP $z$ axis (the vector orthogonal to the plane of the FIELDS antennas) is pointed off of the Sun and rotates with a period of 24 minutes about the sun-spacecraft line, tracing out a conical shape with the body of the spacecraft.

\begin{table}
\centering
\begin{tabular}{ccccc}
    \hline
    \multirow{2}{*}{Year} & \multirow{2}{*}{Month} & Start & Stop\\
    & & (DD HH:MM) & (DD HH:MM)\\
    \hline\hline
    2020 & Dec & 03 12:00 & 03 17:00\\
    \hline
    \multirow{2}{*}{2020} & \multirow{2}{*}{Apr} & 23 01:30 & 23 10:00\\
    & & 23 12:00 & 24 00:00\\
    \hline
    \multirow{2}{*}{2020} & \multirow{2}{*}{Mar} & 14 06:30 & 14 14:00\\
    & & 14 15:00 & 14 18:00\\
    \hline
    \multirow{2}{*}{2019} & \multirow{2}{*}{Jul} & 21 01:00 & 21 08:00\\
    & & 21 10:00 & 22 07:00\\
    \hline
    \multirow{2}{*}{2018} & \multirow{2}{*}{Dec} & 17 12:10 & 18 02:00\\
    & & 18 03:30 & 18 12:00 \\
    \hline
\end{tabular}
\caption{Time intervals during coning roll maneuvers omitting transient Jovian or solar emission. Start and stop times are given in UTC. Adapted from P22.}
\label{tab:roll_maneuvers}
\end{table}

\begin{figure}
    \centering
    \includegraphics[width=\columnwidth]{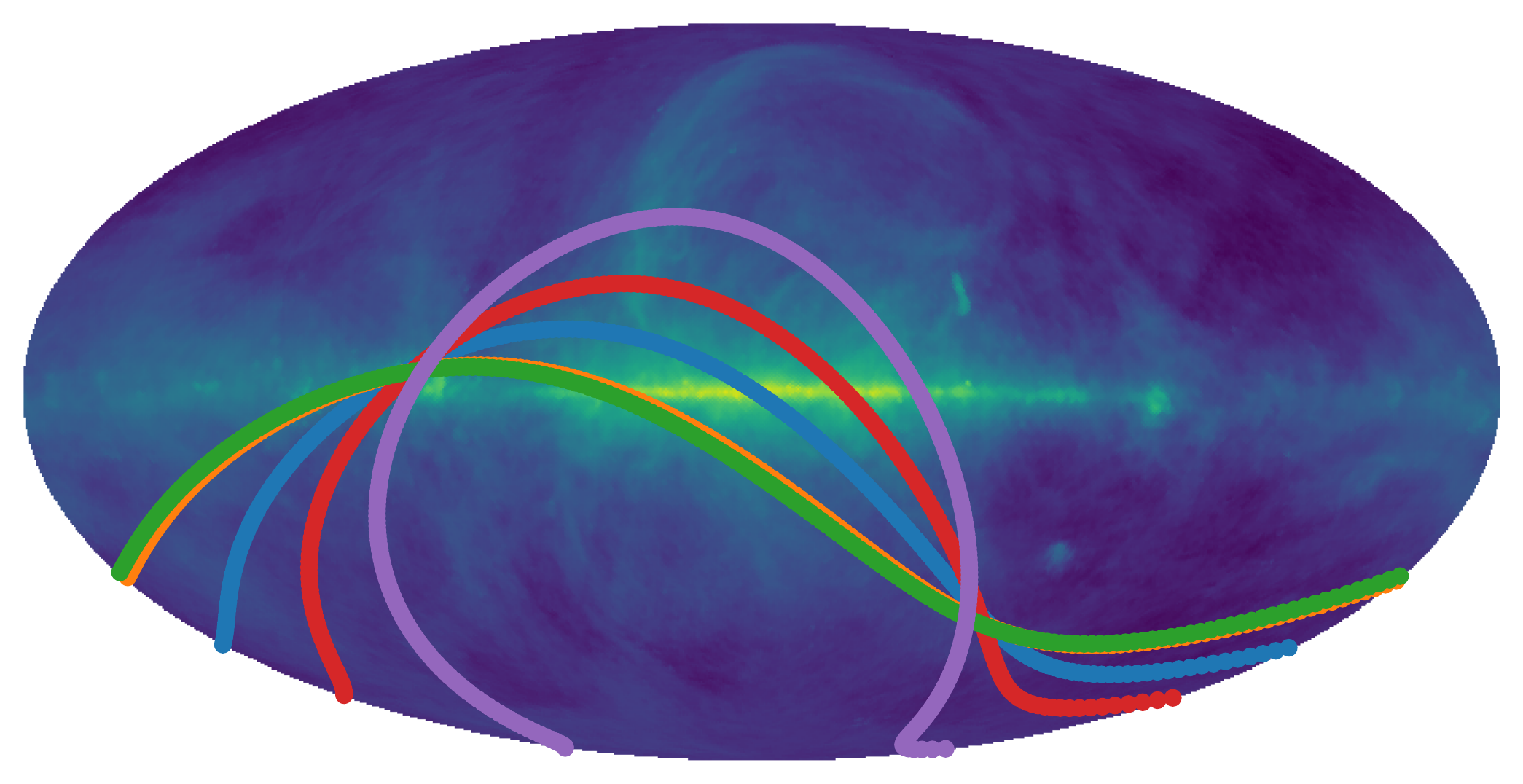}
    \caption{Pointings of the FIELDS V1 antenna during each of the five coning roll maneuvers: 12/03/2020 (blue), 04/23/2020 (orange), 03/14/2020 (red), 07/21/2019 (green), 12/17/2018 (purple). The curves are plotted over the \cite{Haslam:1982} 408 MHz map to provide a sense of the orientations of the antennas relative to the galaxy.}
    \label{fig:coning_roll_pointings}
\end{figure}

Though each coning roll maneuver generally lasts for between 10 and 24 hours, transients such as solar bursts or Jovian emission events can contaminate observations. In an attempt to limit contamination from these transients, we excise data surrounding the events by eye. While the excision limits the data to be analyzed, increasing the noise level and thus potentially reduces the constraining power, our goal is to restrict the FIELDS observations to the sky emission as much as possible. Table \ref{tab:roll_maneuvers} indicates the time intervals during roll maneuvers that were deemed to be clean of enhanced Jovian or solar emission. Since the five coning rolls that we will analyze are spread over a period of nearly two years and multiple PSP orbits, the orientation of the spacecraft relative to the galaxy during the roll changes each time. The pointing of the V1 antenna over the course of a full rotation period during each of the five coning roll maneuvers is shown in Figure~\ref{fig:coning_roll_pointings}.

Autocorrelation spectra $\langle VV^{*}\rangle$, i.e. power spectral density, with units nV$^2$/Hz are calculated for each effective dipole. Though the antenna configuration also permits the calculation of cross-correlations between the two effective dipoles, we leave analysis of these data to future work\footnote{P22 incorporates cross-correlation data in the spherical harmonic decomposition analysis.}. $\langle VV^{*}\rangle$ can be straightforwardly converted to brightness $B_{\nu}$ (with units $\textup{W}\ \textup{m}^{-2}\ \textup{Hz}^{-1}\ \textup{sr}^{-1}$) through the relation
\begin{equation}
\label{eq:brightness_conversion}
    \langle VV^{*}\rangle = \frac{4\pi}{3}Z_0\Gamma^2l_{\textup{eff}}^2B_{\nu},
\end{equation}
where $Z_0 = \sqrt{\mu_0 / \epsilon_0}$ is the impedance of vacuum, $\Gamma$ is the gain factor, and $l_{\textup{eff}}$ is the effective length of the dipole \citep{Zaslavsky:2011}. Lab measurements of a model FIELDS antenna \citep{Pulupa:2017} indicate that $\Gamma \approx 0.32$.
P22 found that $l_{\textup{eff,V1-V2}} = 3.3 \pm 0.1$~m provided the best agreement with the spectrum published in \citealt{Novaco:1978}. For the results presented in this work, a constant value of $l_{\textup{eff,V1-V2}} = 3.3$ m is assumed. Given $l_{\textup{eff,V1-V2}}$, $l_{\textup{eff,V3-V4}}$ can be determined by the ratio of the V1-V2 and V3-V4 autocorrelations. Again following P22, we adopt $l_{\textup{eff,V3-V4}} / l_{\textup{eff,V1-V2}} = 0.99 \pm 0.01$.

Brightness temperature is also a commonly used quantity to characterize low frequency radiation. To convert from power spectral density to brightness temperature $T_b$, the relation is
\begin{equation}
\label{eq:brightness_temperature_conversion}
    \langle VV^{*}\rangle = \frac{8\pi}{3}Z_0k_B\Gamma^2 \bigg(\frac{l_{\textup{eff}}}{\lambda}\bigg)^2 T_b,
\end{equation}
where $k_B$ is Boltzmann's constant and $\lambda$ is the wavelength. The models that we will use are evaluated in brightness temperature before being converted to power spectral density to fit the data.

\subsection{Systematic Noise Subtraction}

After the autocorrelation spectrum is formed, but before any analysis is performed, two known sources of systematic noise are subtracted from the FIELDS spectra: electronic noise from the instrument receiver and QTN from the local plasma environment. Based on pre-flight testing, noise of the form
\begin{align}
\begin{split}
    \Big(7.4 + 0.38(\nu / 1\ \textup{MHz})^{-2}\Big)\bigg(\frac{T_{\textup{PA1}} + T_{\textup{PA2}}}{298\ K}\bigg) +\\
    \Big(6.0\Big)\bigg(\frac{T_{\textup{DCB}}}{298\ K}\bigg)\ \textup{nV}^2\textup{/Hz},
\end{split}
\end{align}
where $T_{\textup{PA1}}$, $T_{\textup{PA2}}$, and $T_{\textup{DCB}}$ are the temperature of pre-amp 1, pre-amp 2, and the digital control board, respectively, is subtracted from the V1-V2 autospectrum. For the V3-V4 autospectrum, the pre-amp temperatures for antennas V3 and V4 are used. The QTN is modeled with a spectrum of the form
\begin{equation}
    A\bigg(\frac{\nu}{1\textup{ MHz}}\bigg)^{-b}\ \textup{nV}^2\textup{/Hz}.
\end{equation}
After the electronic noise is subtracted from the spectrum, $A$ and $b$ are fit using the 300 - 400 kHz region of the FIELD Low Frequency Receiver (LFR) band, where QTN is dominant. These best-fit values are then used to subtract the QTN spectrum from the HFR band.

\subsection{Roll-averaged Spectra}
\label{data_roll_average}

\begin{figure}
    \centering
    \includegraphics[width=\columnwidth]{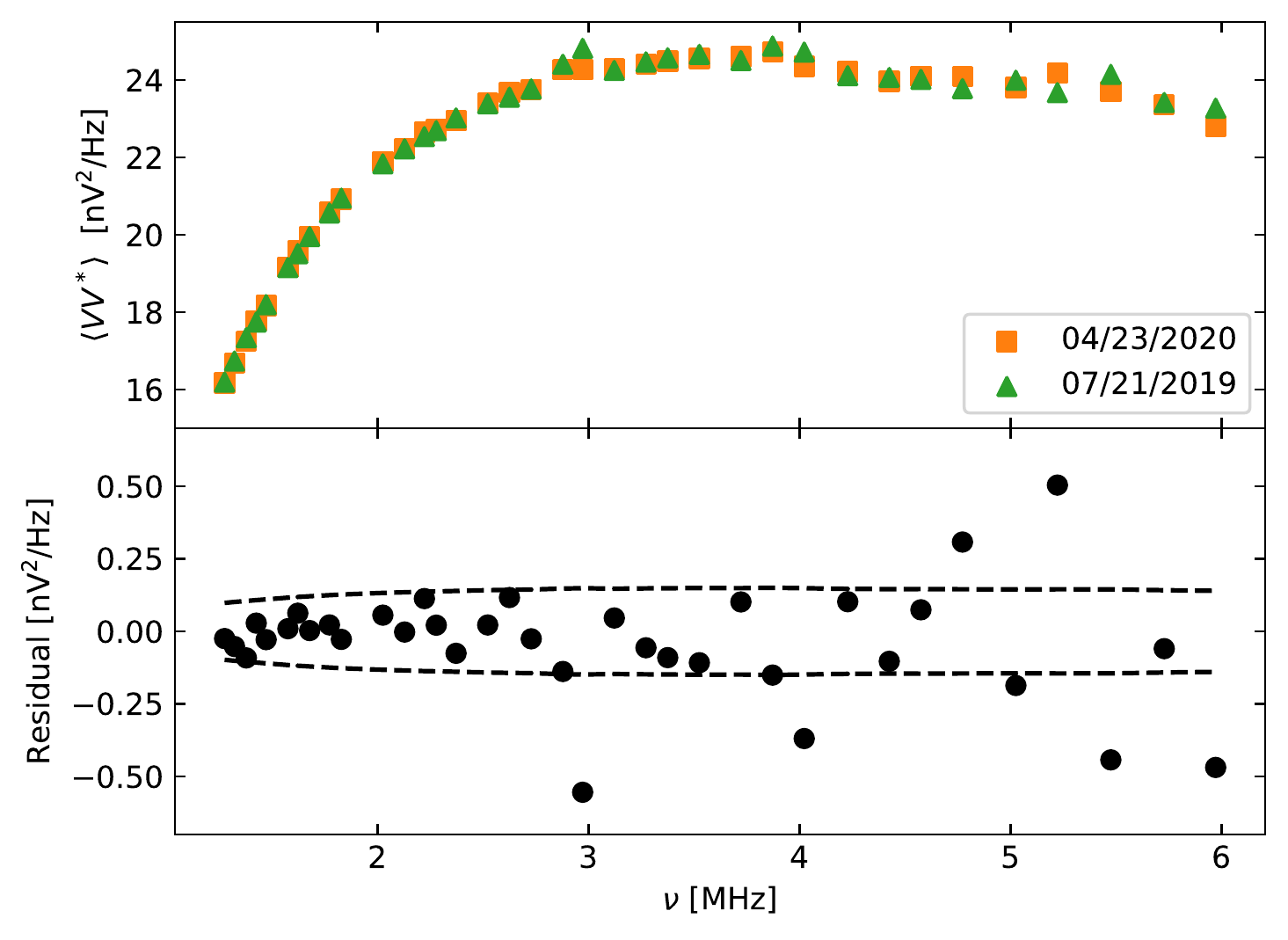}
    \caption{\textit{Top}: roll-averaged spectra for the 04/23/2020 and 07/21/2019 coning rolls. \textit{Bottom}: residual between the two spectra plotted in the top panel. The dashed line indicates the assumed error level for a roll averaged spectrum consisting of both systematic and statistical components.}
    \label{fig:roll_average_comparison}
\end{figure}

In order to perform the analysis, we assemble the FIELDS data into two different formats: roll-averaged and phase-binned. Averaging the data over the course of each roll maneuver produces a set of five spectra, one for each of the rolls indicated in Table \ref{tab:roll_maneuvers}. Note that the roll-averaged component differs slightly from the all-sky monopole component (i.e. the $l = 0$, $m = 0$ spherical harmonic) due to the fact that the FIELDS antennas do not ``see'' the entire sky with equal sensitivity over the course of a roll. The roll-averaged data are most useful in assessing the spectral behavior of the model, particularly the frequency at which free-free absorption produces a maximum in brightness.

We take the error on a single auto spectrum $\langle VV^{*}\rangle$ to be
\begin{equation}
    \sigma_{\langle VV^{*}\rangle} = \frac{\langle VV^{*}\rangle}{\sqrt{240}}.
\end{equation}
The factor of $240 = 80 \times 3$ comes from the fact that 80 spectra recorded over an interval of 2 seconds along with 3 adjacent frequency channels are averaged together on board the spacecraft to produce the spectrum that is ultimately telemetered to the ground. Many of these raw auto spectra are averaged together to obtain the roll-averaged spectrum.
The statistical error on the roll-averaged mean is
\begin{equation}
\label{eq:statistical_error}
    \sigma_{\textup{roll-average,statistical}} = \frac{\langle VV^{*}\rangle}{\sqrt{240 \times n_{\textup{spectra}}}},
\end{equation}
where $n_{\textup{spectra}}$ is the number of spectra that have been averaged together. In general, using data from the time periods specified in Table \ref{tab:roll_maneuvers}, $n_{\textup{spectra}} \approx 5,000$, meaning that the normalized statistical error approaches 0.1\% or smaller.



Figure~\ref{fig:roll_average_comparison} compares the roll-averaged spectra from the 04/23/2020 and 07/21/2019 roll maneuvers. As shown in Figure~\ref{fig:coning_roll_pointings}, the orientations of the FIELDS antennas during these rolls are very similar. As expected, the two roll-averaged spectra are nearly identical with the exception of some scatter that appears to be uncorrelated in frequency. Since the synchrotron emission that makes up most of the sky is expected to be spectrally smooth, as well as the fact that the spectra are derived from very similar views of the sky, it is unlikely that the scatter is intrinsic to the emission. The apparently uncorrelated nature of the scatter also seems unlikely to be caused by transient events that have escaped our cleaning process. The statistical error (calculated with Equation~\ref{eq:statistical_error}) is not large enough to account for the observed scatter, so it must be caused by some source of systematic error. We can still account for this error in our analysis without absolute knowledge of its source. Fortunately, since the scatter appears to be uncorrelated in frequency there will be little overlap between this systematic error and the models we will be fitting to the data. We can straightforwardly estimate the magnitude of the error and include it as a diagonal term in the covariance matrix. We estimate this systematic error as a 0.6\% fractional error. The total error (statistical + systematic) is plotted in Figure~\ref{fig:roll_averaged_reconstruction} relative to the spectral residuals.


To fit the five roll-averaged spectra simultaneously, we concatenate the spectra into a single data vector $\boldsymbol{y}$\footnote{Throughout this paper we will use bold symbols to refer to vectors or matrices.} with 5 ($n_{\textup{rolls}}$) $\times$ 36 ($n_{\textup{frequencies}}$) = 180 channels. We also construct a covariance matrix $\boldsymbol{C}$ which accounts for all sources of error, both statistical and systematic. $\boldsymbol{C}$ is a square matrix with shape $180 \times 180$, with
\begin{equation}
\label{eq:covariance_matrix}
    C_{ij} = \begin{cases}
        \textup{Var}\big[y_i\big] & i = j\\
        \textup{Cov}\big[y_i, y_j\big] & i \neq j
    \end{cases},
\end{equation}
where $C_{ij}$ is the element in the $i$th row and $j$th column of $\boldsymbol{C}$. In this case, since both the statistical and systematic error are assumed to be uncorrelated, the covariance matrix is diagonal with all off-diagonal elements equal to 0.
This covariance matrix is used in the likelihood function to perform the nonlinear fits described in Section \ref{sec:fitting}.

\subsection{Phase-binned Spectra}

\begin{figure}
    \centering
    \includegraphics[width=\columnwidth]{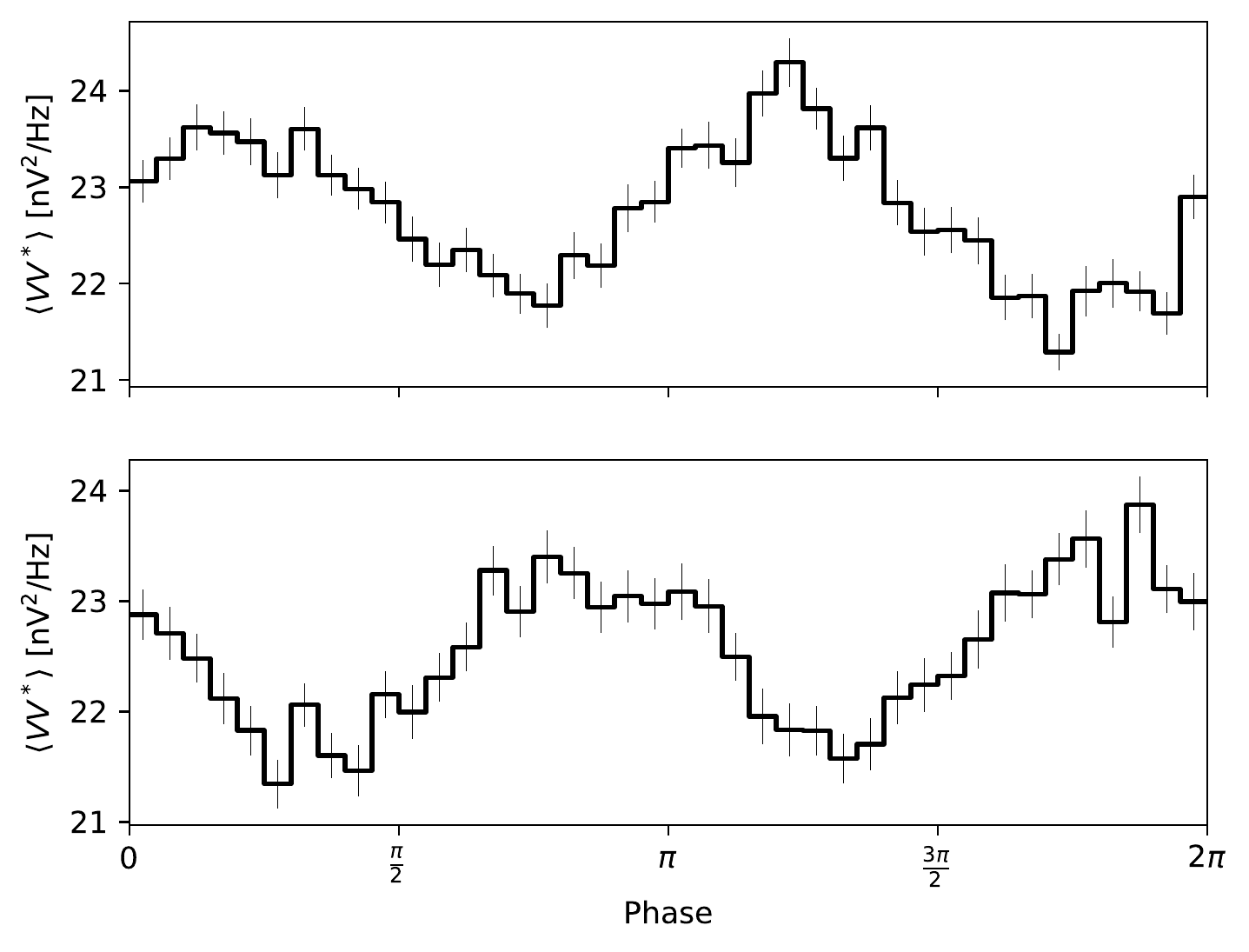}
    \caption{Phase-binned V1-V2 (top) and V3-V4 (bottom) autocorrelations at 5.97 MHz using 40 phase bins during the April 23, 2020 coning roll maneuver. Error bars show the $1\sigma$ uncertainty calculated from Equation~\ref{eq:phase_bin_error}.}
    \label{fig:phase_bin_example}
\end{figure}

To more clearly illustrate the modulation in the FIELDS observations induced by the PSP coning roll maneuvers, we phase-fold and bin the RFS spectra at each frequency. While the phase-folded data are derived from the same observations as the roll-averaged data, we find it useful to inspect each separately in order to better assess the performance of the model. The roll-averaged spectra can be thought of as a single phase bin, while increasing the number of bins breaks the roll into increasingly smaller sections.

Errors for each phase bin at each frequency are given by the standard error of the bin mean, i.e.
\begin{equation}
\label{eq:phase_bin_error}
    \sigma_{\textup{bin}} = \frac{\sigma}{\sqrt{N}},
\end{equation}
where $N$ is the number of data points within the bin and $\sigma$ is their standard deviation. Figure~\ref{fig:phase_bin_example} shows an example of the phase-binned data for a single frequency channel. In this example, V3-V4 appears to lag V1-V2 by a phase of about $\pi / 2$ or so, which makes sense given the orientation of the antennas. Comparing the magnitude of the peak-to-trough distance to the average magnitude of the phase-binned data indicates that the roll-induced modulations are about 10\% or less of the average sky magnitude. Although only a single frequency channel is shown in this plot, the phase-binned data can be concatenated into a single data vector with shape $2 \times n_{\textup{bins}} \times n_{\textup{frequencies}}$, allowing all frequencies and both autocorrelations to be fit simultaneously. The covariance matrix is formed using the square of the error given by Equation~\ref{eq:phase_bin_error} for the diagonal elements. Again $\boldsymbol{C}$ is diagonal with off-diagonal elements equal to 0.

\section{Modeling and Fits}
\label{modeling}

Our forward model of the FIELDS observations consists of three main components: the antenna beam, which quantifies the sensitivity of the antennas to incoming radiation; an emissivity function, which provides the inherent brightness of the sky along a given line of sight; and an absorption model, which describes how the emission is attenuated by free-free absorption. The synthetic model makes use of the SPICE toolkit \citep{Acton:1996, Acton:2018} to obtain the attitude of the PSP spacecraft over the course of a coning roll maneuver. SPICE kernel data provide the pointing of the FIELDS antennas in galactic coordinates, which is used to orient the antenna beams relative to the emission and absorption maps. Since each coning roll involves a periodic movement of the antennas, to make the synthetic model we retrieve the orientations of the antennas from the SPICE toolkit at the phase bin centers during a single roll period and use these orientations to construct simulated observations. This model is then fit to either the roll-averaged or phase-binned observations as described in Section~\ref{results}.

\subsection{Antennas}
\label{antennas}

We employed the CST\footnote{\url{https://www.3ds.com/products-services/simulia/products/cst-studio-suite/}} electromagnetic simulation software to compute the farfield radiation pattern of PSP/FIELDS' four antennas. The spacecraft and antenna 3D models were constructed in CST based on general dimensions extracted from the publicly available PSP's 3D CAD model\footnote{\url{https://solarsystem.nasa.gov/system/resources/usdz_files/2356_PSP.usdz}}. Since the antennas operate at the low frequency (i.e., large wavelength) regime, the antenna beam patterns are primarily affected by the physical size of the major components on the spacecraft. Only five major components are included in the simplified CST model: four antennas, a front heatshield, a hexagonal payload, two radiators, and two main solar panels. The material properties of those components (Table~\ref{tab:psp_model_material}) were also simplified to be either lossy\footnote{Accounted for skin depth effect.} aluminum or idealized insulation foam.

\begin{figure*}
    \centering
    \subfloat[PSP's publicly available 3D CAD model.]{\includegraphics[width=0.45\textwidth]{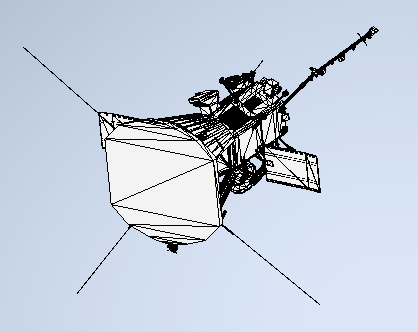}}
    \subfloat[Overlay of the 3D V1-V2 dipole beam at 5.025 MHz on the simplified PSP model in CST.]{\includegraphics[width=0.5\textwidth]{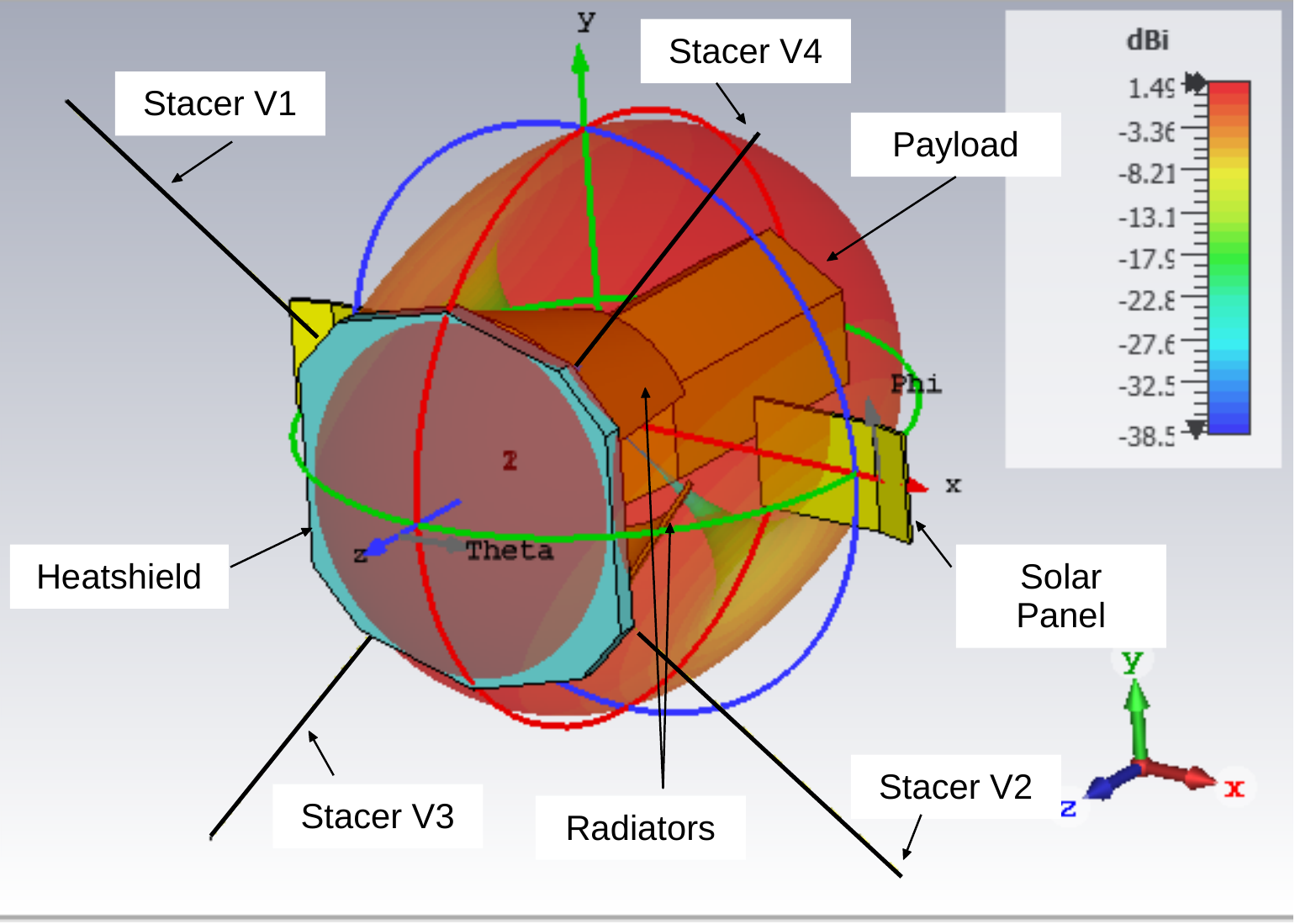}}
    \caption{3D rendered models of the PSP spacecraft and major components included in the CST electromagnetic simulation. Note the commonly recognizable toroidal beam pattern for a dipole in free space.}
    \label{fig:psp_spacecraft}
\end{figure*}

\begin{figure*}
    \centering
    \subfloat[E plane beam cut.]{\includegraphics[width=0.5\textwidth]{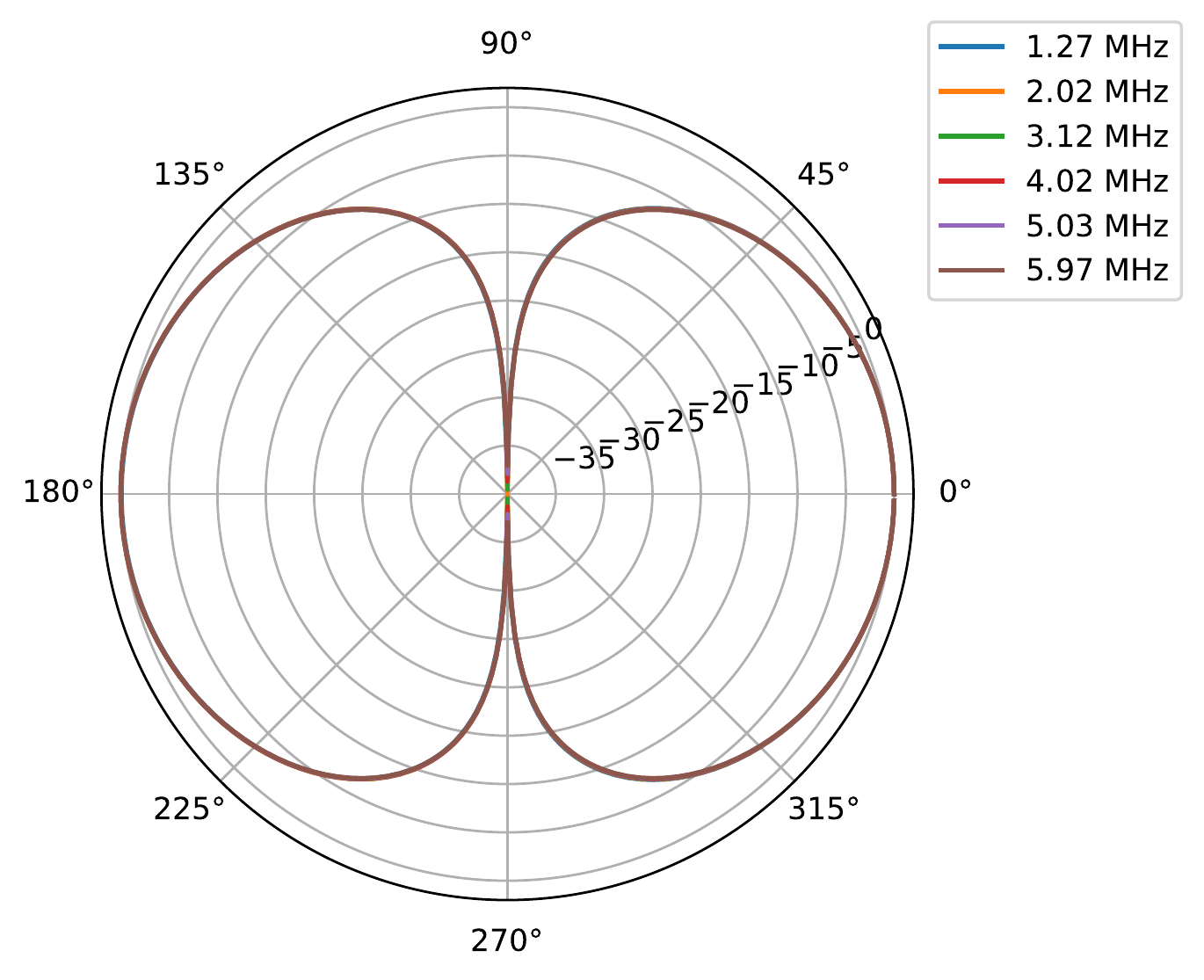}}
    \subfloat[H plane beam cut.]{\includegraphics[width=0.5\textwidth]{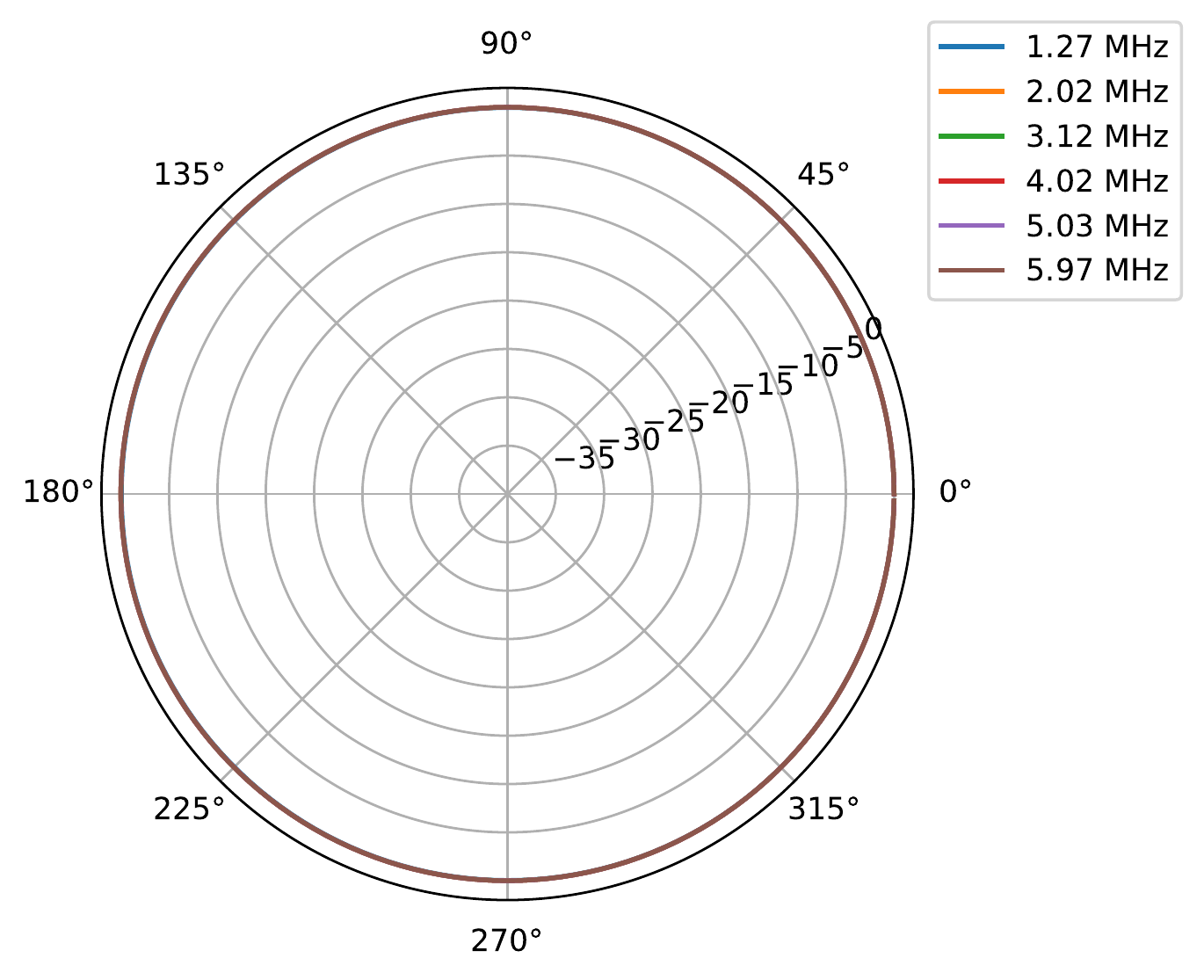}}
    \caption{Polar plots of two different cuts through the FIELDS Stokes I antenna beams from numerical simulations performed with CST Microwave Studio software for a subset of 6 frequency channels from the FIELDS RFS band. The beams are plotted in dB and are normalized such that the maximum value is 0. There is little variation in the normalized Stokes I beams with frequency, causing the curves to overlap significantly.}
    \label{fig:FIELDS_beams}
\end{figure*}

Each of the four antennas was simulated in situ separately, using the CST's 3D full-wave\footnote{The 3D full-wave solver approximates solutions for the complete set of Maxwell's equations without any simplifying assumption, such as 2D quasi-static.} Time Domain Solver\footnote{The Time Domain Solver utilizes the Finite Integration Techniques solver method. The PSP model was simulated with open perfectly match layer boundary conditions in all six directions of the simulation domain box, resulting in a total meshcell number of 18,203,328.}, for each frequency channel between 1 and 6 MHz. The farfield beam patterns of the two oppositely located monopole antennas were combined coherently with $180^{\circ}$ phase offset to produce the toroidal free-space dipole beam pattern (shown in Figures~\ref{fig:psp_spacecraft} and \ref{fig:FIELDS_beams}). For this work, we calculate only the total power Stokes I beam, though the crossed-dipole system does permit Stokes Q, U, and V polarization beams to be calculated. The equations governing the convolution of the beam pattern with the sky are given in Section \ref{synthetic_observations}.

\begin{table}
    \setlength{\tabcolsep}{6pt}
    \centering
    \begin{tabular}{ll}
       \hline
       Component & Assigned Material \\
       \hline\hline
       Antennas & Lossy Aluminum  \\
       Heatshield & Foam ($\epsilon_r = 1$) \\
       Hexagonal Payload & Lossy Aluminum   \\
       Radiators & Lossy Aluminum  \\
       Solar Panels & Lossy Aluminum \\
       \hline
    \end{tabular}
    \caption{Assumed material properties for the simplified PSP components used in the CST electromagnetic simulation.}
    \label{tab:psp_model_material}
\end{table}

\subsection{Emission}
\label{emission}

Previous efforts to model galactic emission, particularly above $\sim$40 MHz (such as the GSM mentioned above), rely on the interpolation of partial or full sky maps from the measured frequency to the frequency of interest. However, for lower frequencies, interpolation poses several potential problems. First, there is no measured sky map below the FIELDS band with which to anchor the interpolation from higher to lower frequencies. Second, interpolation does not offer a straightforward way to parameterize the model such that it can be varied in order to fit the FIELDS observations.

Instead, we adopt an analytic function for the emissivity of the galaxy in a galactocentric cylindrical coordinate system,
\begin{equation}
\label{eq:emissivity}
    \varepsilon(\nu, R, Z) = A\bigg(\frac{R + r_1}{R_0}\bigg)e^{-R/R_0}e^{-|Z/Z_0|}\bigg(\frac{\nu}{\nu_0}\bigg)^{\beta}\ \textup{K/kpc},
\end{equation}
where $r_1 = 0.1$ kpc is an offset to prevent a value of 0 occurring at the origin and $\nu_0 = 408$ MHz is the reference frequency for which the spectral index $\beta$ is defined. This is a modified version of the function used in the ULSA model\footnote{The ULSA emissivity function contains two additional parameters $\alpha$ and $\gamma$, which were omitted here due to their strong covariance with $R_0$ and $Z_0$. Tests with the ULSA emissivity function indicated that these additional parameters did not improve the goodness-of-fit.} \citep{Cong:2021} and describes an axisymmetric emissivity that falls off exponentially in the radial and $Z$ directions, with scale heights $R_0$ and $Z_0$, respectively. $A$, $R_0$, $Z_0$, and $\beta$ are treated as free parameters, while $r_1$ and $\nu_0$ are fixed.

In addition to galactic emission, there is also an extragalactic component of emission from unresolved sources that is approximately isotropic (before accounting for free-free absorption). Again, we follow \cite{Cong:2021} and adopt
\begin{equation}
    T_E(\nu) = 1.2\bigg(\frac{\nu}{1\ \textup{GHz}}\bigg)^{-2.58}\ \textup{K}
\end{equation}
as the brightness temperature of the isotropic extragalactic emission. This emission will be attenuated by the same free-free absorption that affects galactic emission, as described in Section \ref{absorption}.

\subsection{Absorption}
\label{absorption}

At frequencies below $\sim$10 MHz, free-free absorption results in a significant attenuation of incident radiation. The optical depth of the absorption can be approximated by
\begin{equation}
    \tau = 3.28 \times 10^{-7}\bigg(\frac{T_e}{10^4\ \textup{K}}\bigg)^{-1.35}\bigg(\frac{\nu}{\textup{GHz}}\bigg)^{-2.1}\bigg(\frac{\textup{EM}}{\textup{pc cm}^{-6}}\bigg),
\end{equation}
where $T_e$ is the electron temperature and EM is the emission measure \citep{Condon&Ransom:2016}. Since the absorption will be dominated by free electrons in the warm interstellar medium (WIM), we use $T_e = 8000$ K. EM is given by the integral of the squared electron density along a line-of-sight (LOS), i.e.
\begin{equation}
    \textup{EM} = \int n_e^2 ds.
\end{equation}

Several different models of the galactic free electron distribution exist, namely NE2001 \citep{Cordes&Lazio:2002} and YMW16 \citep{Yao:2017}. Both models derive estimates of $n_e$ from the measured dispersion of pulse arrival times from known pulsars. For this work we will use the YMW16 model of the free electron density due to its ability to better estimate pulsar distances, particularly at high galactic latitudes. Complicating the estimation of EM from the model, however, is that pulsar measurements can only be used to calculate the dispersion measure (DM), given by
\begin{equation}
    \textup{DM} = \int n_e ds.
\end{equation}
If the free electron density is constant along the LOS, then DM can be converted directly to EM. Unfortunately, free electrons are generally clumped into clouds, such that the free electron density is not constant. Suppose (as an illustrative example) that the free electron density along a LOS can be described by $n_e(s) = \overline{n_e} + \delta_{n_e}(s)$, where $\delta_{n_e}(s)$ is a fluctuation on the average density $\overline{n_e}$. Assuming that the fluctuations average to 0 over a given LOS, then $\textup{DM} = \int \overline{n_e}ds$ and thus the fluctuation term can effectively be ignored. However, these fluctuations cannot be ignored when calculating EM, which gives $\textup{EM} = \int(\overline{n_e} + \delta_{n_e})^2ds = \int \overline{n_e}^2ds + \int \delta_{n_e}^2ds$.\footnote{Note that a third term with a single power of $\delta_{n_e}$ on the right-hand-side of this equation is 0 because of the assumption that the fluctuations average to 0 over the line of sight.}

A detailed derivation of the EM (see P22 Section 5), taking into account the clumping of free electrons into discrete clouds as well as intra-cloud density fluctuations, provides an expression for the EM calculated from the synthetic density $\overline{n}_e$, provided by the pulsar-based models:
\begin{subequations}
\begin{align}
        \textup{EM} &= \int f^{-1}\overline{n}_e^2ds,\\
        f &= \big(\eta^{-1}\zeta + Fl_0^{2/3}\big)^{-1}.
\end{align}
\end{subequations}
In the above expression, $\eta$ is the fraction of the LOS occupied by clouds, $\zeta = \langle n_e^2\rangle / \langle n_e\rangle^2$ quantifies the inter-cloud variations in density\footnote{Brackets indicate an average over the portion of the LOS filled by clouds.}, $F = \eta^{-1}\zeta \epsilon^2 l_0^{-2/3}$ is the so-called ``fluctuation parameter,'' and $\epsilon^2 = {\delta n_e^2}/n_e^2$ describes the intra-cloud variations. This expression assumes that the intra-cloud fluctuations follow a power law relationship in wavenumber with index 11/3 between an inner scale $l_1$ and an outer scale $l_0 \gg l_1$. Generally, $l_0 \approx 1$ pc. For most portions of the galaxy, $f$ is dominated by the inter-cloud variations, i.e. $f \approx \eta \zeta^{-1}$. Although sometimes density fluctuations within clouds can become significant (e.g. near the galactic center), we will generally refer to $f$ as the ``filling factor'' due to its dependence on $\eta$, the fraction of the LOS filled with clouds.

Unfortunately, the filling factor is relatively unconstrained, making conversion from DM to EM difficult. In this paper, we divide the filling factor into four galactic components: the thick disk, the thin disk, the spiral arms, and the galactic center (GC). \cite{Yao:2017} models the radial dependence of electron density in the thin disk, spiral arms, and galactic center using sech$^2$ functions that peak at each component's central radius and fall off with some characteristic scale length.
For this work, a point is deemed to be ``within'' each component (for the purposes of the filling factor) if it is within two scale lengths of the center of the component, as given by the relevant equations of \cite{Yao:2017}. In cases where the components overlap with each other, we use the filling factor of the component that comes first in the list [galactic center, spiral arms, thin disk]. For these three components, the filling factor is assumed to be constant.

The rest of the galactic volume is assumed to be the thick disk. Motivated by \cite{gaensler:2008}, which found that the the filling factor increases exponentially from the galactic plane with a scale height of approximately 0.7 kpc, we assume that the filling factor is dependent on the $z$ galactic coordinate. The authors of \cite{gaensler:2008} also suggest that above $\sim$2 kpc, the filling factor decreases until $\sim$5 kpc, where it reaches a constant value. We take a functional form for the filling factor of the thick disk that increases exponentially from a value $a$ in the plane to $ae^2$ at two scale heights above the plane and subsequently decreases exponentially to an asymptotic value $b$ at high latitudes, i.e.
\begin{equation}
    f = \begin{cases}
        a e^{|z| / 0.7} & |z| < 1.4\ \textup{kpc}\\
        b\big[1 - \big(1 - \frac{ae^2}{b}\big)e^{-(|z| - 1.4) / 0.7}\big] & |z| \geq 1.4\ \textup{kpc}\\
    \end{cases}.
\end{equation}

Simply due to its larger volume relative to the thin disk, spiral arms, or galactic center, the thick disk and its filling factor are more important in terms of the magnitude of the free-free absorption compared to the other galactic components. Specifically $a$, which determines the filling factor near the plane where the free electron density is highest, is likely to be the most important quantity with regards to the absorption. $b$, which describes the filling factor at high latitudes, where the free electron density is much lower, has a much smaller impact on the absorption. In order to minimize the dimensionality of the parameter space as much as possible, which increases the efficiency of nonlinear sampling methods, we treat only $a$ as a free parameter, while holding $b = 0.01$, $f_{\textup{thin}}^{-1} = 7$, $f_{\textup{arm}}^{-1} = 3$, and $f_{\textup{GC}}^{-1} = 1 \times 10^5$ at constant values. These values are consistent with those used in P22.


\subsection{The Sun}

\begin{table}
    \setlength{\tabcolsep}{4pt}
    \centering
    \begin{tabular}{ccccc}
       \hline
       date & $d_{\odot}$ (AU) & $D_{\odot}$ ($^{\prime}$) & $\Omega_{\odot}$ (sr) & ($l$, $b$)$_{\odot}$ \\
       \hline\hline
       12/03/2020 & 0.795 & 40.2 & 1.076e-4 & (48.26, -53.69)\\
       04/03/2020 & 0.815 & 39.3 & 1.024e-4 & (68.95, -60.86)\\
       03/14/2020 & 0.816 & 39.2 & 1.023e-4 & (34.92, -44.56)\\
       07/21/2019 & 0.809 & 39.6 & 1.041e-4 & (71.34, -61.23)\\
       12/17/2018 & 0.801 & 40.0 & 1.061e-4 & (23.56, -30.68)\\
       \hline
    \end{tabular}
    \caption{Quantities related to the Sun relative to PSP for each of the five coning roll days. The quantities are the distance between the Sun and PSP, the apparent angular diameter of the Sun in arcminutes (the angular diameter of the Sun as viewed from Earth is approximately 30$^{\prime}$), the solid angle subtended by the Sun, and the apparent galactic longitude and latitude coordinates of the Sun.}
    \label{tab:sun_size}
\end{table}

Assuming that any transient solar emission, such as flares or bursts, have been removed through the excision process described above, the quiescent Sun contributes a brightness temperature of approximately 5800 K, the blackbody temperature of the solar photosphere. The brightness temperature of the Sun is dwarfed by that of the galaxy, but the Sun also acts as a blocking body, occulting the galaxy behind it. Table \ref{tab:sun_size} provides the distance, apparent angular diameter, and apparent galactic coordinates of the Sun relative to PSP. The angular sizes of the Sun in Table \ref{tab:sun_size} assume the Sun ends at the photosphere. While plasma at higher solar radii may still obscure the sky at these frequencies (the plasma frequency at 10 solar radii can be $\sim$1 MHz), even increasing the apparent angular size of the sun by an order of magnitude would still result in an extremely small effect, as described in the following paragraph.
While the Sun is slightly larger for PSP compared to an observer on Earth, its angular size is still relatively small due to the fact that the coning rolls are performed near the apocenter of PSP's orbit.

The Sun will have the biggest effect on the sky brightness if it is blocking the brightest portion of the sky. Assuming that the portion of sky blocked by the Sun is constant at this maximum brightness $T_{\textup{max}}$, then the fractional portion of the all-sky brightness blocked by the Sun is
\begin{equation}
    \frac{\Omega_{\odot}}{4\pi}\frac{T_{\textup{max}}}{T_{\textup{mean}}}.
\end{equation}
Given the solid angles in Table \ref{tab:sun_size}, $\Omega_{\odot} / 4\pi \approx 10^{-5}$. Estimating the $T_{\textup{max}} / T_{\textup{mean}}$ is a bit more difficult since low resolution maps will tend to underestimate this ratio. For the high resolution ($n_{\textup{side}}$ = 128) Haslam map at 408 MHz \citep{Haslam:1982}, the ratio is $T_{\textup{max}} / T_{\textup{mean}} \approx 20$. The max to mean ratio is unlikely to be larger than this in the FIELDS band because the absorption will likely reduce the maximum apparent brightness. Using $T_{\textup{max}} / T_{\textup{mean}} = 20$, the fractional difference in the all-sky brightness caused by the Sun is (at worst) of order $10^{-4}$, or a hundredth of a percent. While differences of this order may be important for extremely sensitive analyses, such as extraction of the 21-cm signal, it is dwarfed by both the statistical and systematic uncertainties of the FIELDS observations. We do implement the presence of the Sun in our modeling (using the quantities in Table \ref{tab:sun_size}), but it is not a significant component of the sky brightness and its contribution to any uncertainty in the model can be considered negligible.

\subsection{Synthetic Observations}
\label{synthetic_observations}

This section describes how each of the components described above are used to construct the synthetic FIELDS observations that make up the model that is fit to the data. The sky viewed by the antennas (before accounting for the beam pattern) is given by an integral along the LOS:
\begin{align}
\begin{split}
\label{eq:T_sky}
    T_{\textup{sky}}(l, b, \nu) = &\int_0^{s_g}\varepsilon(l, b, s, \nu)e^{-\tau(l, b, s, \nu)}ds\\
    &+ T_E(\nu)e^{-\tau(l, b, s_g, \nu)},
\end{split}
\end{align}
where $s_g$ is the pathlength to the edge of the galaxy. The beam of either the V1-V2 or V3-V4 effective dipole is then rotated to the correct position on the sky according to the SPICE orientation data, and the brightness temperature of the synthetic observation is given by
\begin{equation}
\label{eq:T_ant}
    T_{\textup{ant}}(\nu) = \frac{\int_{4\pi}B(\theta, \phi, \nu)T_{\textup{sky}}(\theta, \phi, \nu)d\Omega}{\int_{4\pi}B(\theta, \phi, \nu)d\Omega}.
\end{equation}
The Stokes I beam function $B(\theta, \phi, \nu)$, describing the sensitivity of the antennas to incoming unpolarized radiation, is provided by the CST simulation described in Section \ref{antennas}.\footnote{Though a small amount of polarized radiation can potentially couple into the Stokes I spectrum through ``polarization leakage,'' we do not account for this effect here.}

The synthetic antenna temperature spectrum is calculated for 500 evenly spaced points over the course of a single roll period. The antenna temperature in Kelvin is then converted to power spectral density through Equation~\ref{eq:brightness_temperature_conversion}. For fitting the roll-averaged spectra, these 500 synthetic spectra are averaged together, whereas for the phase-binned data a cubic spline interpolation is performed in order to evaluate the model at the center of the phase bins.

Although the above equations that govern real observations involve continuous integrals, in order to calculate the synthetic observations, we must rely on discrete sums that estimate the true values of the integrals. We utilize the \texttt{HEALPix} formalism \citep{Gorski:2005} to discretize maps of the antenna beam, absorption, and emission. Each time the parameters of the absorption or emission change we must re-calculate the integrals in Equations \ref{eq:T_sky} and \ref{eq:T_ant}. For each pixel, we sample $s$ in steps of 0.1 kpc out to a maximum distance of 50 kpc. Each time we wish to create a synthetic observation, we must evaluate both the emission and absorption model at each step in $s$ for each map pixel for each frequency channel. In order to perform a nonlinear fit, which involves constructing potentially hundreds of thousands of synthetic observations, we use $n_{\textup{side}} = 4$, which corresponds to $n_{\textup{pixels}} = 3,072$. This results in relatively low spatial resolution maps, but is mitigated by the extremely large size of the beam. Higher values for $n_{\textup{side}}$ result in intractable computation times to perform a fit, even with high performance computing resources.

\subsection{Fitting}
\label{sec:fitting}

Let $\boldsymbol{y}$ be the data vector of the five concatenated roll-averaged spectra and let $\boldsymbol{\mathcal{M}}(\boldsymbol{\theta})$ be the synthetic model of the data with parameters $\boldsymbol{\theta}$, which is constructed from the components described above. To fit the model to the data we construct a likelihood function
\begin{align}
\begin{split}
    \mathcal{L}(\boldsymbol{y} | \boldsymbol{\theta}) = &|2\pi\boldsymbol{C}|^{-1/2}\\
    &\times \textup{exp}\bigg\{-\frac{1}{2}[\boldsymbol{y} - \boldsymbol{\mathcal{M}}(\boldsymbol{\theta})]^T\boldsymbol{C}^{-1}[\boldsymbol{y} - \boldsymbol{\mathcal{M}}(\boldsymbol{\theta})]\bigg\},
\end{split}
\end{align}
where $\boldsymbol{C}$ is the covariance matrix described in Equation \ref{eq:covariance_matrix}. The data vector, covariance matrix, and the model are derived from either the roll-averaged data or the phase-binned data, which are fit separately, though the parameters of the model are the same. Ultimately, we want to explore the posterior distribution of the parameters given the data, $p(\boldsymbol{\theta} | \boldsymbol{y})$, which is given by Bayes' theorem,
\begin{equation}
\label{eq:bayes}
    p(\boldsymbol{\theta} | \boldsymbol{y}) = \frac{\mathcal{L}(\boldsymbol{y} | \boldsymbol{\theta}) \cdot \pi(\boldsymbol{\theta})}{\mathcal{Z}},
\end{equation}
where $\pi(\boldsymbol{\theta})$ is the prior distribution on the parameters and the normalization term $\mathcal{Z}$ is often referred to as the evidence.

To sample from the posterior distribution and perform parameter estimation, we use a nested sampling algorithm, which is able to efficiently sample parameter spaces with multi-modal distributions or significant degeneracies between parameters.\footnote{Nest Sampling is better in this regard compared to Markov Chain Monte Carlo (MCMC) algorithms, another commonly used sampling method.} Specifically, we use the \texttt{PyMultiNest}\footnote{\url{https://github.com/JohannesBuchner/PyMultiNest}} implementation \citep{Buchner:2014} of the \texttt{MultiNest} algorithm \citep{Feroz:2009, Feroz:2019}.

\begin{table}
    \centering
    \begin{tabular}{ccc}
        \hline
        Parameter & Units & Prior \\
        \hline\hline
        $a$ & -- & Unif(0, 1)\\
        \hline
        $A$ & K kpc$^{-1}$ & Unif(0, 200)\\
        \hline
        $R_0$ & kpc & Unif(0, 30)\\
        \hline
        $Z_0$ & kpc & Unif(0, 30)\\
        \hline
        $\beta$ & -- & Unif(-3, -2)\\
        \hline
    \end{tabular}
    \caption{Parameters of the model $\boldsymbol{\mathcal{M}}(\boldsymbol{\theta})$ for both the roll-averaged and phase-binned data and their prior distributions.}
    \label{tab:priors}
\end{table}

\begin{figure}
    \centering\includegraphics[width=\columnwidth]{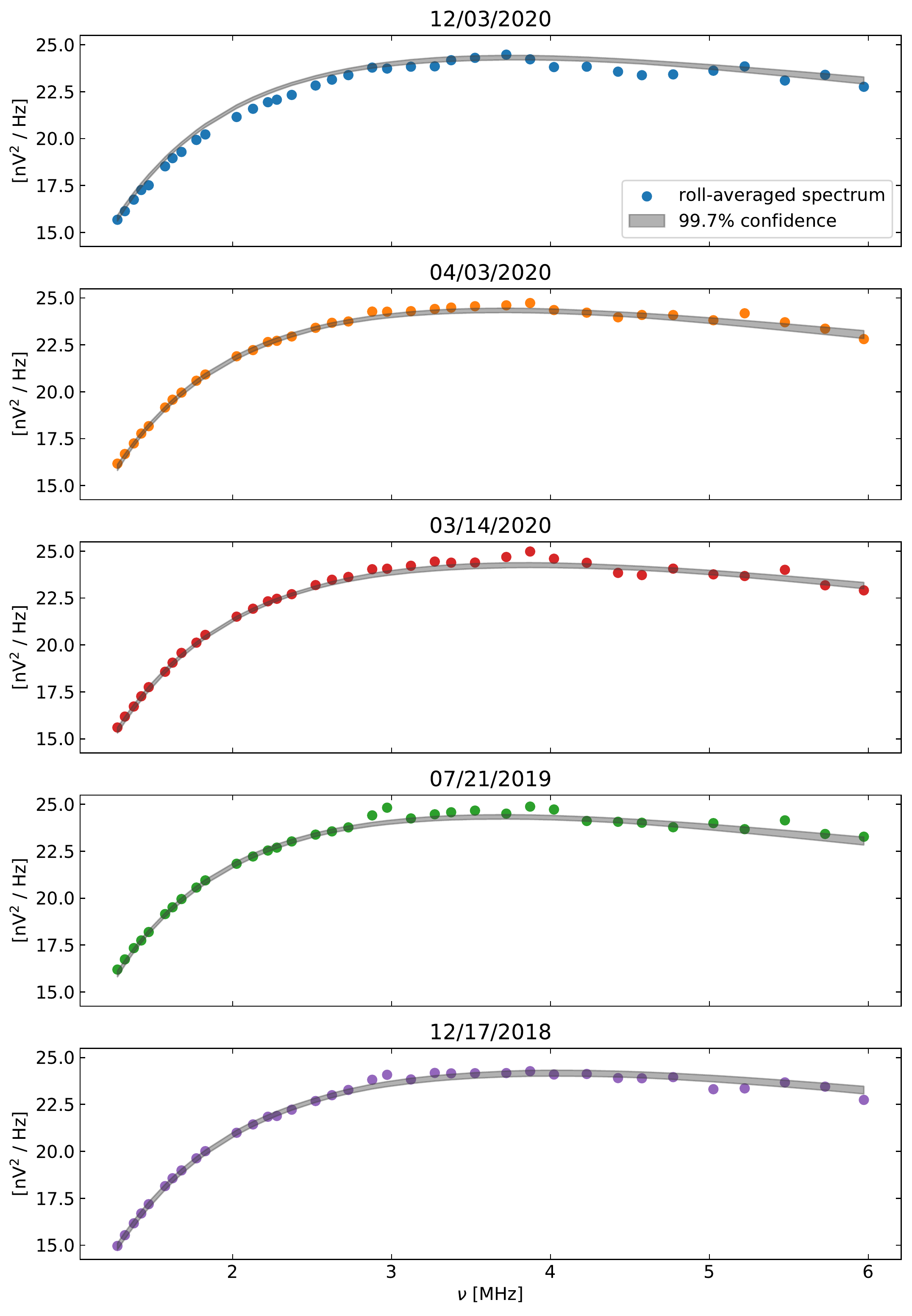}
    \caption{FIELDS HFR roll-averaged spectra for the five coning roll maneuvers listed in Table \ref{tab:roll_maneuvers} (solid dots). The gray contours indicate the 99.7\% confidence intervals obtained by fitting the simulated observation model to the FIELDS observations using \texttt{PyMultiNest}.}
    \label{fig:roll_averaged_reconstruction}
\end{figure}

The parameters of the model and their prior distributions that are given to \texttt{PyMultiNest} to begin the sampling are shown in Table \ref{tab:priors}. 
$a$ describes the filling factor of the absorption in the thick-disk component at the galactic plane (see Section \ref{absorption}), while $A$, $R_0$, $Z_0$, and $\beta$ describe the magnitude, radial scale height, vertical scale height, and spectral index (in brightness temperature) of the emission (see Section \ref{emission}). The primary hyper-parameters of the nested sampling algorithm are $n_{\textup{live}}$ and $\mathcal{Z}_{\textup{tol}}$, which describe the number of active samples utilized by the algorithm and the tolerance on the Bayesian evidence, which is used as a stopping criterion. For the results presented in the following section, we use $n_{\textup{live}} = 5000$ and $\mathcal{Z}_{\textup{tol}} = 0.1$.\footnote{While the default value for $n_{\textup{live}}$ is 400, we found that we needed to increase this value significantly to obtain accurate results, likely due to the degeneracies present in the parameter space.}


\section{Results and Discussion}
\label{results}

\subsection{Roll-averaged Spectra}
\label{sec:fitting_roll_averaged_spectra}

\begin{figure*}
    \centering
    \includegraphics[width=0.95\textwidth]{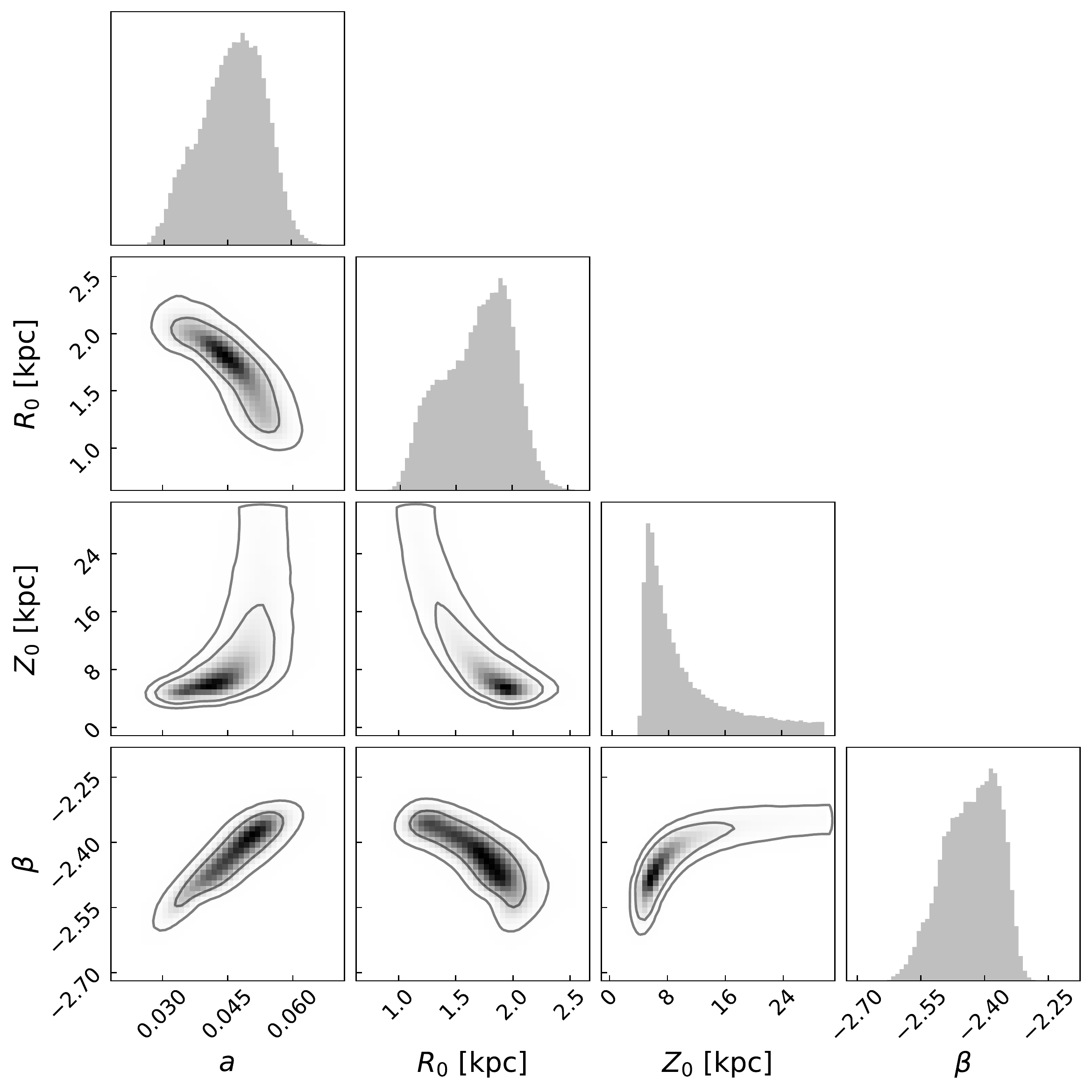}
    \caption{Corner plot of the posterior distribution of 4 model parameters for a fit to the roll-averaged FIELDS data. While $A$ (see Equation~\ref{eq:emissivity}) is fit as a free parameter, we marginalize over it. Data from all 5 roll maneuvers are fit simultaneously. The two contours in each panel show the 68 and 95\% confidence areas, while the shaded grid squares plot the two-dimensional histogram for each pair of parameters. The one-dimensional histograms show the full marginalized posterior distribution for each parameter.}
    \label{fig:roll_averaged_triangle}
\end{figure*}

We first fit the roll-averaged spectra, fitting all five rolls simultaneously. A quantitative assessment of the validity of fitting data from different roll maneuvers simultaneously is contained in Appendix \ref{appendix_tension}. Figure~\ref{fig:roll_averaged_reconstruction} compares the 99.7\% confidence interval of the best-fit reconstruction to the FIELDS roll-averaged spectra for each of the 5 coning roll maneuvers. Examining the reconstruction intervals, we find that they agree quite well with the FIELDS spectra.
In each of the five spectra, the model is able to reproduce the maximum in the brightness between 3 and 4 MHz as well as the slope both above and below the maximum. As discussed in Section \ref{data_roll_average}, there is noticeable scatter, particularly above 3 MHz, but this is unlikely to be intrinsic to the sky and is accounted for by the covariance matrix used to perform the fit. Since we are fitting a nonlinear model, we choose not to use the traditional chi squared goodness-of-fit statistic because of the difficulty in estimating the number of degrees of freedom for nonlinear models (see, e.g., \citealt{Andrae:2010}). Instead, we examine the distribution of the normalized residuals in Appendix \ref{appendix_residuals}.

\begin{table}
    \centering
    \begin{tabular}{ccc}
        \hline
        Parameter & Units & MAP Estimate \\
        \hline\hline
        $a$ & -- & $0.046^{+ 0.007}_{- 0.009}$\\
        \hline
        $R_0$ & kpc & $1.746^{+ 0.232}_{- 0.390}$\\
        \hline
        $Z_0$ & kpc & $7.775^{+ 9.064}_{- 2.831}$\\
        \hline
        $\beta$ & -- & $-2.431^{+ 0.065}_{- 0.085}$\\
        \hline
    \end{tabular}
    \caption{One-dimensional maximum a posteriori estimates of the four non-marginalized free parameters from the roll-averaged fit. Uncertainties on the MAP estimates encompass the 68\% confidence interval.}
    \label{tab:1D_parameter_distributions}
\end{table}

While the model reconstructions are contained in a small interval shown by the gray contours in Figure~\ref{fig:roll_averaged_reconstruction}, the parameter values can still vary significantly and there exist strong covariances between some of the parameters. Figure \ref{fig:roll_averaged_triangle} displays the parameter posterior distribution in detail through a corner plot, which plots the posterior of 4 of the 5 free parameters. The fifth free parameter, $A$, has been marginalized because of its direct degeneracy with the $\Gamma$ and $l_{\textup{eff}}$ calibration parameters (see Equation~\ref{eq:brightness_conversion}) and lack of physical relevance compared to the other free parameters. The most notable covariances agree with general expectations of how the model should behave. For example, the spectral index $\beta$ and the thick disk filling factor $a$ have a strong positive covariance. As $a$ increases, the free-free optical depth decreases, leading to less absorption, which must be balanced by a more positive spectral index.

One-dimensional constraints on each parameter are given in Table \ref{tab:1D_parameter_distributions}. The values of each of the parameters are mostly in agreement with expectations. \cite{gaensler:2008} estimated $a = 0.04 \pm 0.01$, which is consistent with the constraint of $a = 0.046^{+ 0.007}_{- 0.009}$ obtained from our fit.
$R_0 = 1.746^{+ 0.232}_{- 0.390}$ and $Z_0 = 7.775^{+ 9.064}_{- 2.831}$ appear to be reasonable given that the radius of the Milky Way is $\sim 30$ kpc. Note, however, the large upper uncertainty interval for $Z_0$ caused by the tail of the distribution, which extends to the prior bound at $Z_0 = 30$ kpc. Perhaps an important discrepancy in these two parameters is that $Z_0 > R_0$ for the entire posterior distribution, which is inconsistent with \cite{Cong:2021}, which found $R_0 = 3.41$ kpc and $Z_0 = 1.12$ kpc.\footnote{\cite{Cong:2021} did not publish uncertainties for these estimates.} Again, though, these values were obtained by fitting a map that is two orders of magnitude higher in frequency than the FIELDS HFR band. $\beta = -2.5$ is generally used as a fiducial value for the spectral index, which is consistent with our result of $\beta = -2.431^{+ 0.065}_{- 0.085}$.

\subsection{Phase-binned Spectra}

\begin{figure*}
    \centering
    \includegraphics[width=0.75\textwidth]{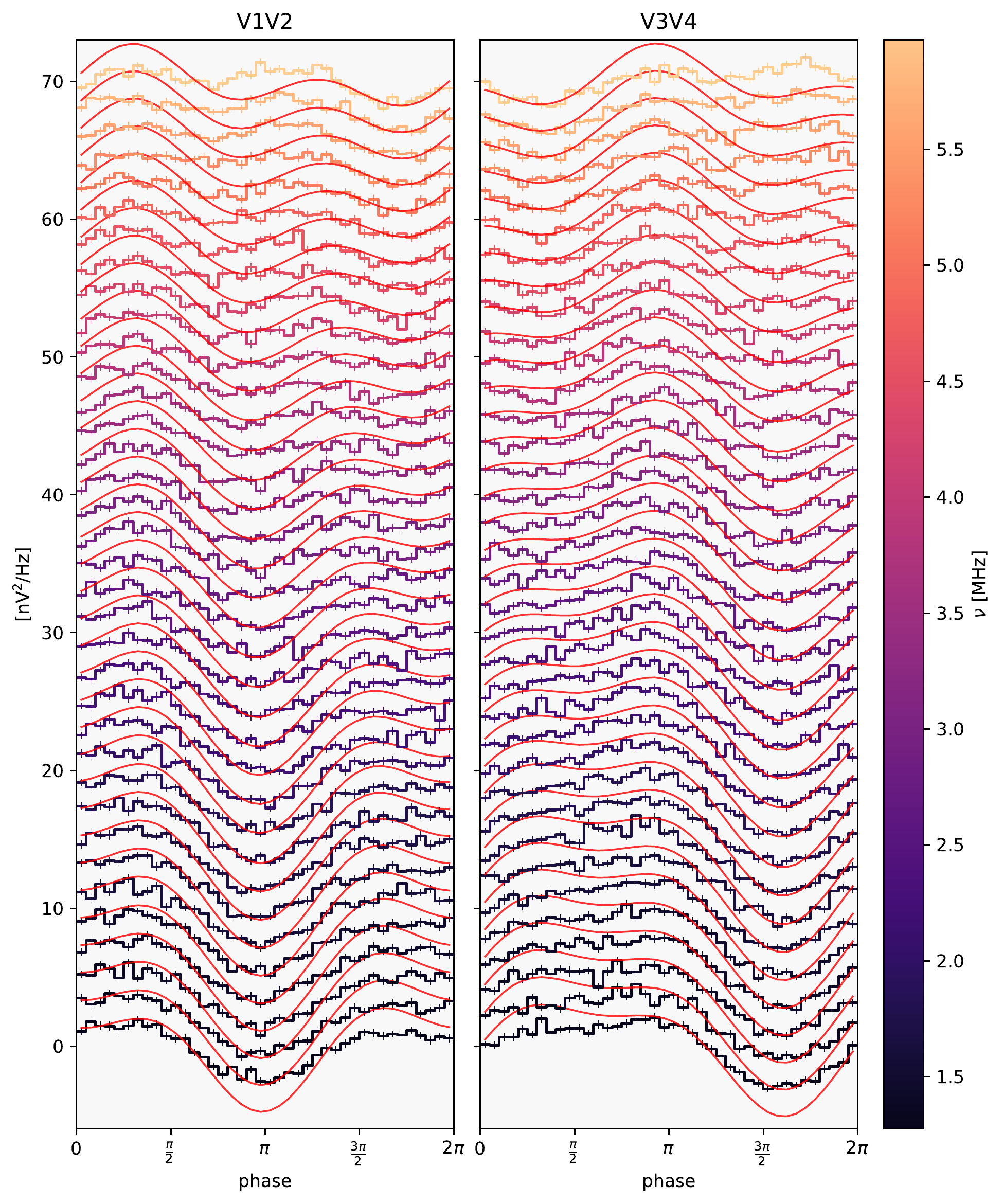}
    \caption{V1-V2 (left) and V3-V4 (right) antenna autocorrelations from the 03/14/2020 roll maneuver binned into 40 equal segments of roll phase compared to a forward model of the FIELDS observations (red curves) using the maximum a posteriori parameter values from the roll-averaged fit (Figure~\ref{fig:roll_averaged_triangle}). Each binned curve is a different frequency channel, with the color of the curve corresponding to the frequency indicated by the color bar on the right. Each binned curve is mean subtracted before an artificial offset of 2 nV$^2$/Hz times the index of the frequency channel is added to separate the curves such that they can be viewed simultaneously. The error bars on each bin are calculated according to Equation~\ref{eq:phase_bin_error}.}
    \label{fig:phase_bin_fit}
\end{figure*}

Now that we have fit the roll-averaged spectra with our model, we turn to the phase-binned data. If a model is an accurate representation of the true sky, the same parameter values should be able to fit both the roll-averaged and phase-binned data, since both are derived from the same observations viewing the same sky. To this end, in Figure~\ref{fig:phase_bin_fit} we compare our forward model of the phase-binned observations using the maximum a posterior (MAP) values of the roll-averaged fit from Figure~\ref{fig:roll_averaged_triangle} to the phase-binned data from the 03/14/2020 roll maneuver. A visual assessment demonstrates a clear mismatch between the data and the model. Even though the peaks and troughs of both the model and the data appear to coincide in phase, their magnitudes are generally quite different. We show data from only one roll maneuver in Figure~\ref{fig:phase_bin_fit}, but comparisons to data from the other four roll maneuvers yield similar levels of (dis)agreement.

If instead of using the roll-averaged MAP parameters for the forward model, we fit the phase-binned data itself, we find that the model generally appears to be insufficient to fully fit the phase-binned data. Utilizing the same nested sampling algorithm that was described in Section \ref{sec:fitting}, we find that the posterior distributions of the model parameters tend to converge towards the edge of the prior boundaries, even when the prior bounds are increased to nonphysical regions of parameter space. For example, $a$, the parameter describing the filling factor of free electrons in the thick disk, prefers values $>1$, which corresponds to greater than 100\% of the line-of-sight being filled with ionized hydrogen, clearly an unrealistic scenario. For this reason, we do not reproduce any posterior distributions or fits to the phase-binned data in this work, instead we argue that the model, particularly the analytic function for the galactic emissivity, is insufficient to accurately describe the phase-binned data. In the following section we describe aspects of the model that can potentially be improved in the future to better represent the true spatial distribution of the sky.

\section{Conclusions}

The purpose of this work is to utilize PSP/FIELDS observations to investigate the low frequency sky between 1 and 6 MHz. While P22, a companion to this paper, uses a model-agnostic spherical harmonic decomposition to describe the spatial distribution of the sky in the $l=0$ and $l=2$ modes, in this work we develop a nonlinear forward model of the observations and fit the model to the FIELDS data using a Bayesian nested sampling algorithm, yielding constraints on the underlying parameters. Before performing the fits, we split the FIELDS data into two different components: the roll-averaged spectrum, in which we average data from the entire roll, and the phase-binned spectrum, in which the roll period is broken into segments of equal phase.

Fitting the roll-averaged data with our five parameter model (one parameter associated with absorption and four parameters associated with emission) produces well-behaved posterior constraints for all five parameters. Comparing the MAP model reconstruction with the roll-averaged spectra shows good agreement. In general, the parameter constraints agree with previously published estimates, with the exception of the $R_0$ and $Z_0$ scale heights of the emissivity function, which prefer slightly larger values than were given in \cite{Cong:2021}.

While our forward model appears to be able to represent well the roll-averaged data, this is not the case for the phase-binned data. Using the MAP parameter values from the roll-averaged fit gives a reconstruction that is in poor agreement with the phase-binned data. If instead we fit the phase-binned data itself using the same nested sampling algorithm, the posterior parameter distributions become bunched near the edge of the prior bounds, attempting to reach unphysical portions of the parameter space. Even after significantly increasing the prior bounds, the model is still unable to provide a good fit. This inability to fit is likely caused by the greater amount of information about the spatial structure of the sky in the phase-binned data compared to the roll-averaged data. Averaging spectra together over the entire rolls smears the structure of the sky together, discarding some information in the process. This smearing effect seems to smooth out the sky to a sufficient level that it can be fit by our model. The phase-binned data, in contrast, does not appear to smooth the sky to a sufficient level, revealing the inadequacy of the model.

There are several ways in which our forward model is likely insufficient and may be improved in the future to better fit the phase-binned FIELDS data (and other future low frequency experiments, such as upcoming lunar-based observations). Perhaps the most obvious shortcoming of the model is that the emissivity function is quite smooth in terms of spatial structure, whereas the true emission at these frequencies likely has significant small scale structure. While the analytic emissivity function used in this work has a significant advantage in terms of ease of evaluation, it cannot accurately represent small-scale structure. The emissivity function is also symmetric between the northern and southern galactic hemispheres. In reality, galactic features such as the northern polar spur and Loop I are likely to break this symmetry. In order to better represent the true spatial structure of the sky, the emissivity function likely needs to be altered in some way to take these two features into account. Finally, the free-free absorption model may also need to be improved. Though the YMW16 model claims to be more accurate than the earlier NE2001 model, estimates of the optical depth are still a large source of potential error. Even with our implementation of the galactic component model for the filling factor, estimating the emission measure, the operative term in the optical depth calculation, is still highly uncertain. Even if the filling factor is known exactly, the model must infer the electron density distribution for the entire galaxy from a sample of only 189 lines-of-sight. A better forward model of the FIELDS observations will likely require a better absorption model, whether that be from a larger number of pulsar dispersion measurements, a more detailed model of galactic structure, or both.

We conclude by noting the implications of this work for future analysis of low frequency observations, particularly efforts to measure the Dark Ages portion of the global 21 cm signal, which lies in the frequency range $\sim 1 - 50$ MHz. Measurement of the global signal involves differentiating it from the emission we have discussed here, which in the case of the Dark Ages can be over 6 orders of magnitudes brighter than the 21 cm signal. While some (such as the EDGES collaboration in \citealt{Bowman:2018}) have utilized general, polynomial-based models to fit the foreground, these methods have the potential to produce false detections (see \citealt{Tauscher:2020b}) even at higher frequencies, where the brightness of the foreground emission relative to the 21 cm signal is smaller. As argued in \cite{paperIII}, the most robust analysis method for extracting the 21 cm global signal involves binning observations according the relative orientation of the antennas on the sky,\footnote{In the case of \cite{paperIII}, the binning is done in local sidereal time (LST) for a ground-based instrument, but this is functionally equivalent to the phase bins used in this work.} taking advantage of the fact that the isotropic global signal remains constant. The results obtained from fitting our forward model of FIELDS observations, and the inability of the model to fit the phase-binned spectra, suggest that significant progress is needed in order to produce a physical model of the sky below 6 MHz that is accurate enough to be used for robust Dark Ages global 21 cm signal extraction.

\section{Acknowledgements}

This work is directly supported by the National Aeronautics and Space Administration (NASA) Solar System Exploration Research Virtual Institute cooperative agreement number 80ARC017M0006. This work was also partially supported by the Universities Space Research Association via DR using internal funds for research development. DR’s work was also partially supported by NASA grant 80NSSC23K0013. This work utilized the Blanca condo computing resource at the University of Colorado Boulder. Blanca is jointly funded by computing users and the University of Colorado Boulder. The National Radio Astronomy Observatory is a facility of the National Science Foundation operated under cooperative agreement by Associated Universities, Inc. Bang~D.~Nhan is a Jansky Fellow of the National Radio Astronomy Observatory. The The Parker Solar Probe/FIELDS experiment was developed and is operated under NASA contract NNN06AA01C.

\bibliographystyle{aasjournal}
\bibliography{ref}

\appendix

\section{Goodness-of-fit of the Roll-averaged Model}
\label{appendix_residuals}

\begin{figure*}
    \centering
    \includegraphics[width=0.75\textwidth]{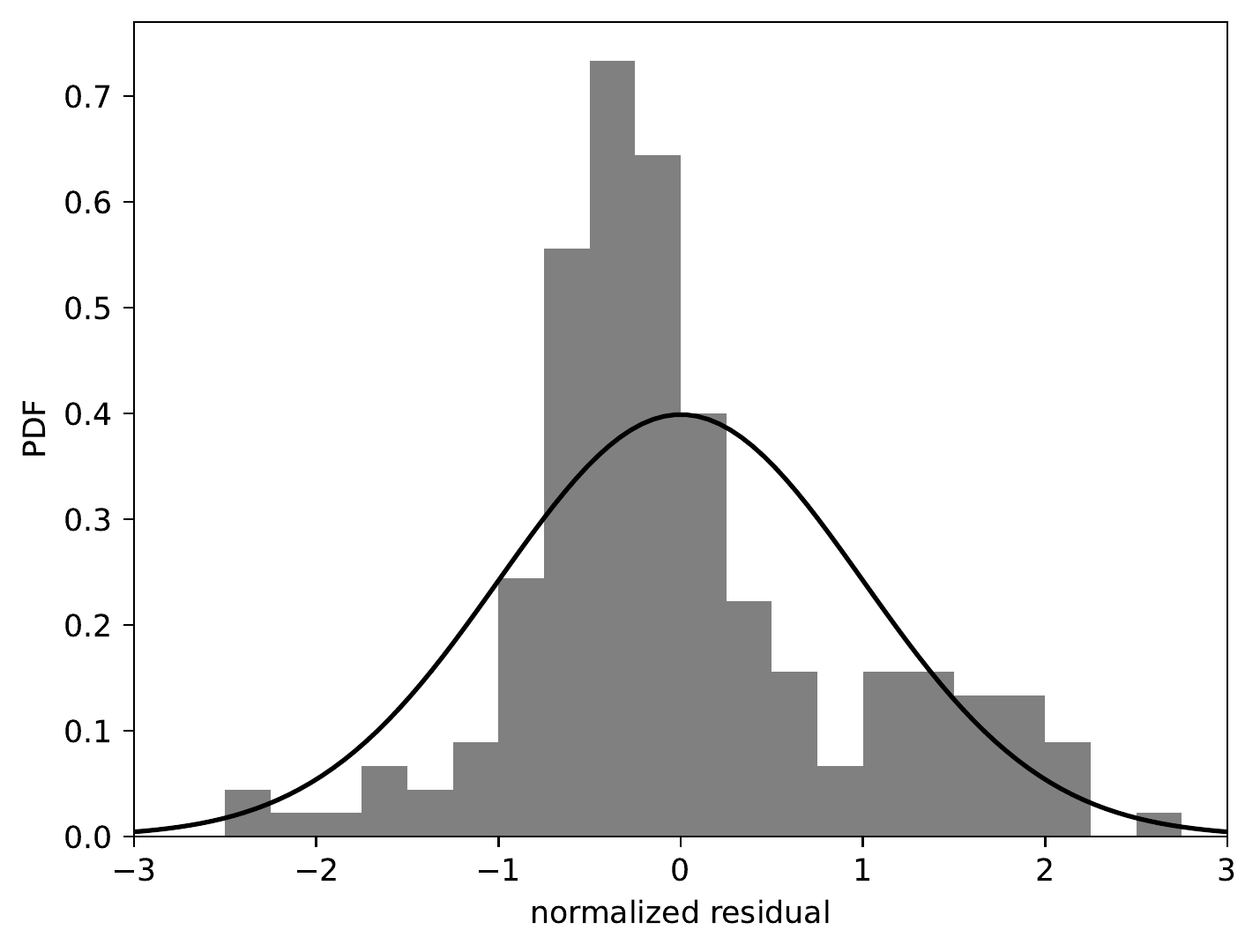}
    \caption{Histogram of the normalized residuals of the best-fit roll-averaged model (grey) compared to a unit Gaussian probability density function (black). While our assumed error appears to have perhaps over-estimated the noise level (indicated by the overdensity in the histogram near 0), using a larger error level ensures that the constraints on the model parameters are not artificially tight.}
    \label{fig:residual_histogram}
\end{figure*}

Calculating the reduced chi squared statistic requires knowledge of the number of degrees of freedom, which cannot be reliably estimated for a nonlinear model. Instead, to assess the goodness-of-fit we follow the suggestion of \cite{Andrae:2010} and examine the residuals of the model fit in detail. Figure \ref{fig:residual_histogram} shows a histogram of the residuals (normalized by the error level) compared to a unit Gaussian distribution. For a model that is able to perfectly represent the data, with true parameter values and known measurement errors, the distribution of the residuals is normal with mean $\mu = 0$ and variance $\sigma^2 = 1$ (i.e. unit Gaussian) by definition.

Inspecting Figure \ref{fig:residual_histogram}, the distribution of the residuals does not appear to conform to a Gaussian. The histogram is more peaked near 0, but if the variance of the Gaussian distribution were reduced, there would be an overdensity in the positive tail. Going back to the assumed error level shown in Figure \ref{fig:roll_average_comparison}, it appears that the error level may be too large for the portion of the frequency band below 2 MHz, leading to the peak in the histogram near 0. While one possible remedy would be to shrink the fractional error level, the error would then likely be underestimated above 2 MHz. The error level could be reduced only in the band below 2 MHz, but there is no \textit{a priori} justification for the fractional error to change with frequency.

Although the normalized residuals may not follow a Gaussian distribution, we note that overestimating the error level is far better than underestimating the error in terms of the model parameter constraints. If the assumed error level of the data is too small, the posterior distribution (i.e., the distribution plotted in Figure \ref{fig:roll_averaged_triangle}) may be too narrow, underestimating the uncertainties on the parameters. For this reason, we deem the error level and the fit sufficient to proceed with our analysis.

\section{Quantifying agreement between roll-averaged spectra}
\label{appendix_tension}

\begin{figure*}
    \centering
    \includegraphics[width=\textwidth]{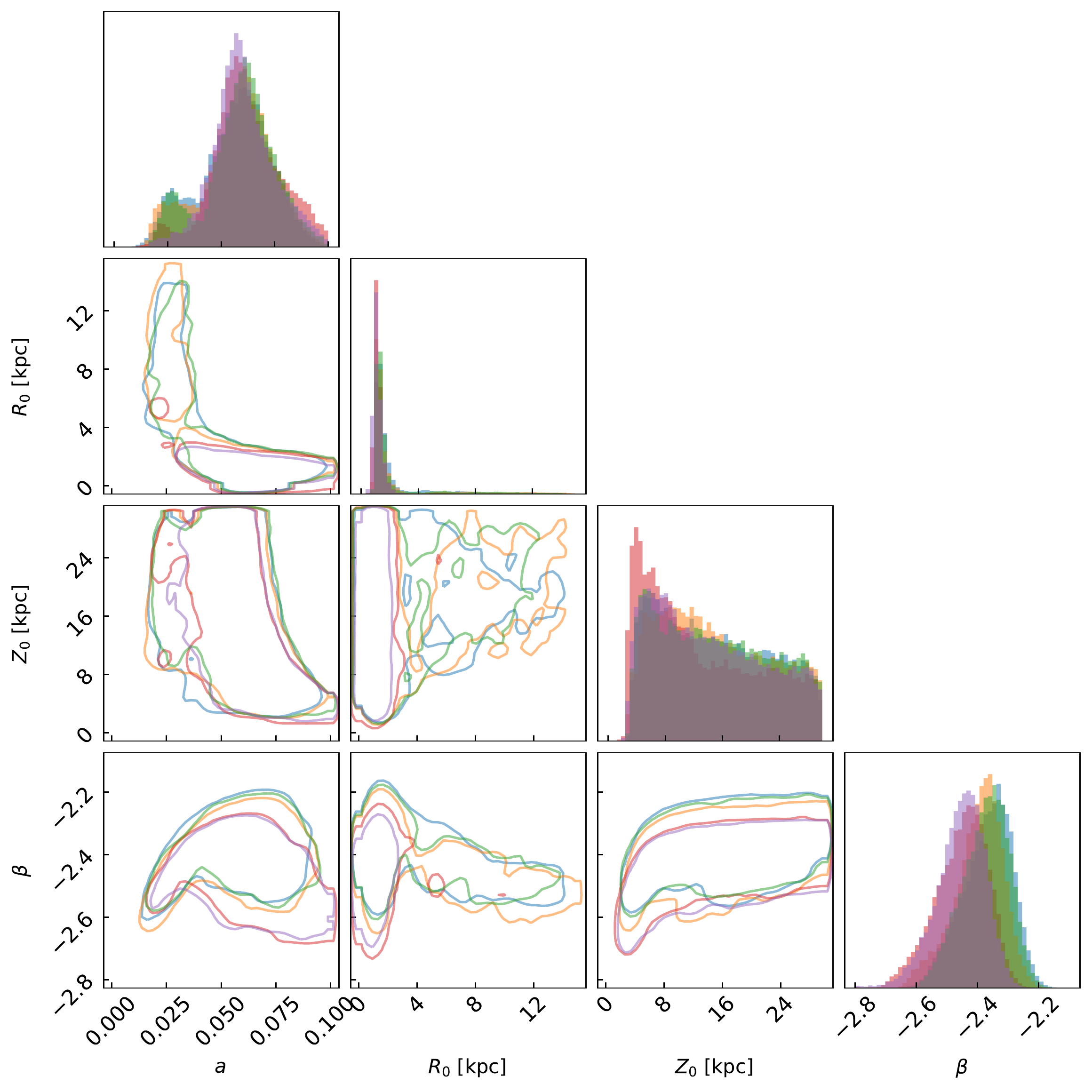}
    \caption{Corner plot of the posterior distribution of 4 model parameters for fits to each of the 5 roll-averaged spectra independently. While $A$ (see Equation~\ref{eq:emissivity}) is fit as a free parameter, we marginalize over it. The contours outline the 95\% confidence region, while the one-dimensional histograms show the full marginalized posterior for each parameter. The colors for each distribution follow the same scheme as in Figure~\ref{fig:coning_roll_pointings}: 12/03/2020 (blue), 04/23/2020 (orange), 03/14/2020 (red), 07/21/2019 (green), 12/17/2018 (purple).}
    \label{fig:triangle_plot_independent}
\end{figure*}

In Section \ref{sec:fitting_roll_averaged_spectra}, we present a joint fit of the five roll maneuvers. In this appendix we demonstrate that jointly fitting the five roll-averaged spectra is justified by demonstrating that the data sets are not in tension under our forward model of the observations. One method (see, e.g., \citealt{Marshall:2006, Grandis:2016}) of quantifying tension between data sets is through the Bayesian evidence. Specifically, one can calculate the evidence ratio
\begin{equation}
    R = \frac{\mathcal{Z}(D_1,D_2)}{\mathcal{Z}(D_1)\mathcal{Z}(D_2)},
\end{equation}
where $D_1$ and $D_2$ are two different data sets and $\mathcal{Z}$ refers to the Bayesian evidence under the chosen model. This definition can be extended to $n$ number of data sets:
\begin{equation}
    R = \frac{\mathcal{Z}(D_1,D_2,\ldots,D_n)}{\prod_{i=1}^n \mathcal{Z}(D_i)}.
\end{equation}
The numerator is the joint evidence of the data sets when fit simultaneously, while the denominator is the product of the evidences of each data set when fit independently. The value of $R$ compares the evidence when both data sets are fit with the same model parameters to the evidence when each data set may be described by a different set of parameters. The value is often interpreted on the so-called Jeffrey's scale \citep{Jeffreys:1961}, with $\ln{R} > 0$ indicating consistency between the data sets and $\ln{R} < 0$ suggesting inconsistency.

While the Jeffrey's scale is commonly used to interpret the value of $R$, particularly in the context of cosmological data sets (see, e.g., \citealt{Raveri:2016}), \cite{Seehars:2016} points out that this scale does not take into account the statistical behavior of $R$, which is a random variable itself. To better account for statistical fluctuations, \cite{Grandis:2016} suggest normalizing the evidence ratio by its variance. Complicating this normalization, however, is the fact that the variance can only be analytically calculated under specific assumptions, most notably that the model is linear. Since our model of the FIELDS observations is highly nonlinear, we are unable to perform the normalization procedure outlined in \cite{Grandis:2016}. However, for reasons that will be explained below, we believe the un-normalized Jeffrey's scale to be sufficient for the analysis performed here.

Figure~\ref{fig:triangle_plot_independent} shows the posterior distributions for each of the five roll-averaged spectra fit independently. Each of the five distributions display significant overlap and the un-normalized evidence ratio value is $\ln{R} = 23.2$. Although we cannot exactly calculate the variance of $R$ under our model, \cite{Grandis:2016} calculate that Var[$\ln{R}$] = $d$/2 for linear models, where $d$ is the number of parameters. In the case of our model, $d = 5$, so the variance in $\ln{R}$ (for a linear model) would be 2.5. Even if the variance of $\ln{R}$ for our nonlinear model were larger by an order of magnitude, we would still conclude that the data sets were consistent to a moderate confidence level. Coupling this with the fact that the five independently-fit posterior distributions visually exhibit significant overlap in the corner plot, we have confidence that the five roll-averaged spectra are consistent to a sufficient level to fit them simultaneously.

\end{document}